\documentclass[lettersize,journal]{IEEEtran}
\usepackage{amsmath,amsfonts}
\usepackage{algorithmic}
\usepackage{algorithm}
\usepackage{array}
\usepackage[caption=false,font=normalsize,labelfont=sf,textfont=sf]{subfig}
\usepackage{textcomp}
\usepackage{stfloats}
\usepackage{url}
\usepackage{verbatim}
\usepackage{graphicx}
\usepackage{cite}
\usepackage[utf8]{inputenc}
\usepackage[T1]{fontenc}

\begin{document}

\title{Positronium Imaging: History, Current Status, and Future Perspectives}

\author{
Pawe{\l} Moskal,
Aleksander Bilewicz,
Manish Das,
Bangyan Huang,~\IEEEmembership{Student Member,~IEEE,}
Aleksander Khreptak,
Szymon Parzych,
Jinyi Qi,~\IEEEmembership{Fellow,~IEEE,}
Axel Rominger,
Robert Seifert,
Sushil Sharma,
Kuangyu Shi,~\IEEEmembership{Member,~IEEE,}
William M. Steinberger,
Rafa{\l} Walczak,
and Ewa St\k{e}pie{\'n}
\thanks{Manuscript received xxx; revised xxx; accepted xxx.
Date of publication xxx; date of current version xxx.}

\thanks{This work was supported by the National Science Centre of Poland through grants MAESTRO no. 2021/42/A/ST2/00423, OPUS no. 2021/43/B/ST2/02150 and OPUS-LAP no. 2022/47/I/NZ7/03112.}

\thanks{(\textit{Corresponding author:} Pawe{\l} Moskal: p.moskal@uj.edu.pl)
\vspace{1em}}

\thanks{
This work did not involve human subjects or animals in its research.
\vspace{1em}}

\thanks{Pawe{\l} Moskal, Manish Das, Aleksander Khreptak, Szymon Parzych, Sushil Sharma and Ewa St\k{e}pie{\'n} are with the Marian Smoluchowski Institute of Physics and the Center for Theranostics, Jagiellonian University, 31-007 Krak{\'o}w, Poland (e-mail: p.moskal@uj.edu.pl; manish.das@doctoral.uj.edu.pl; aleksander.khreptak@uj.edu.pl; szymon.parzych@doctoral.uj.edu.pl; sushil.sharma@uj.edu.pl; e.stepien@uj.edu.pl).}

\thanks{Aleksander Bilewicz and Rafa{\l} Walczak are with the Institute of Nuclear Chemistry and Technology, Warsaw, Poland (e-mail: a.bilewicz@ichtj.waw.pl; r.walczak@ichtj.waw.pl).}

\thanks{Bangyan Huang and Jinyi Qi are with the Department of Biomedical Engineering, University of California at Davis, Davis, CA 95616 USA (e-mail: bybhuang@ucdavis.edu; qi@ucdavis.edu).}

\thanks{Axel Rominger, Robert Seifert, and Kuangyu Shi are with the Department of Nuclear Medicine, Inselspital, Bern University Hospital, University of Bern, 3010, Bern, Switzerland (e-mail: axel.rominger@insel.ch; robert.seifert@unibe.ch; kuangyu.shi@dbmr.unibe.ch).}

\thanks{William M. Steinberger is with Siemens Medical Solutions USA, Inc., Knoxville, TN 37932 USA (e-mail: william.steinberger@siemens-healthineers.com).}
}

\markboth{IEEE Transactions on Radiation and Plasma Medical Sciences,~Vol.~xx, No.~xx, xxx~2025}%
{Shell \MakeLowercase{\textit{et al.}}: A Sample Article Using IEEEtran.cls for IEEE Journals}


\maketitle

\begin{abstract}
Positronium imaging was recently proposed to image the properties of positronium atoms in the patient’s body. Positronium properties depend on the size of intramolecular voids and oxygen concentration; therefore, they deliver information different from the anatomic, morphological, and metabolic images. Thus far, the mean ortho-positronium lifetime imaging has been at the center of research interest. The first \textit{ex vivo} and \textit{in vivo} positronium lifetime images of humans have been demonstrated with the dedicated J-PET scanner, enabling simultaneous registration of annihilation photons and prompt gamma from $\beta^+\gamma$ emitters. Annihilation photons are used to reconstruct the annihilation place and time, while prompt gamma is used to reconstruct the time of positronium formation. This review describes recent achievements in the translation of positronium imaging into clinics. The first measurements of positronium lifetime in humans with commercial PET scanners modernized to register triple coincidences are reported. The \textit{in vivo} observations of differences in ortho-positronium lifetime between tumor and healthy tissues and between different oxygen concentrations are discussed. So far, the positronium lifetime measurements in humans have been completed with clinically available $^{68}$Ga, $^{82}$Rb, and $^{124}$I radionuclides. Status and challenges in developing positronium imaging on a way to a clinically useful procedure are presented and discussed.
\end{abstract}


\newpage

\section{Introduction}

\IEEEPARstart{P}{ositronium} imaging is a newly introduced method that enables imaging of parameters characterizing mechanism of positron-electron annihilation in the living organisms~\cite{Moskal2019a,Moskal2018a,Moskal2021a}. Specifically, \textit{positronium imaging} refers to tomographic imaging of parameters that reflect properties of positronium in the tissue~\cite{Moskal2019b,Moskal2020a}. These are e.g.\ properties of positronium atoms such as (i) mean ortho-positronium (oPs) lifetime ($\tau_{\text{oPs}}$) and (ii) formation probability ($P$), as well as (iii) mean lifetime of positrons undergoing direct annihilations ($\tau_d$), (iv) mean positron lifetime in the tissue ($\tau_{e^+}$), and (v) the 3$\gamma$ to 2$\gamma$ positron annihilation rate ratio ($R_{3/2}$)~\cite{Moskal2021a,Bass2023}. These parameters are sensitive to the molecular environment in which positronium annihilation occurs~\cite{Moskal2019c,Bass2023}. Ortho-positronium lifetime, $\tau_{\text{oPs}}$, depends on the size of free volumes between atoms (voids), and the concentration of paramagnetic molecules such as e.g.\ oxygen as well as on the density of free radicals. The positronium formation probability, $P$, and direct positron-electron annihilations lifetime, $\tau_d$, depend on the free volumes fraction and in general on a spatial arrangement of atoms and their chemical composition. In turn, parameters $\tau_{e^+}$ and $R_{3/2}$ depend on all mentioned factors and are the function of the values of $P$, $\tau_{\text{oPs}}$, and $\tau_d$. Hence imaging of positronium properties in the living organism, and in particular imaging of parameters such as $\tau_{\text{oPs}}$, $\tau_{e^+}$, $\tau_d$, $R_{3/2}$, and $P$, may in principle inform us about the tissue pathology at its early stage of molecular alterations, before the occurrence of functional and morphological changes, and about the degree of hypoxia~\cite{Moskal2019c,Bass2023,Moskal2021b}.

It should be emphasized that \textit{positronium imaging} provides different information to the metabolic imaging (standardized uptake value) and to the anatomic imaging (electron density) as well as to the morphological imaging (hydrogen atom density) achievable by clinical modalities such as positron emission tomography (PET), computed tomography (CT) and magnetic resonance tomography (MR), respectively.

During positron emission tomography, in approximately 40\% of cases, the positron-electron annihilation occurs in the tissue through the formation of a positronium atoms, with approximately three-quarters of positronia formed in the long-lived ortho-positronium state in vacuum which decays almost exclusively to 3$\gamma$~\cite{Bass2023}. In the tissue, the ortho-positronium mean lifetime is substantially shortened and its decay rate to 3$\gamma$ is decreased, since it may also decay to 2$\gamma$ due to the interaction of ortho-positronium with surrounding atoms, paramagnetic molecules and free radicals. This makes ortho-positronium a hallmark of the molecular structure and composition at the nano and sub-nanometre level.

Imaging of the 3$\gamma$ to 2$\gamma$ decay rate ratio can be performed for all kinds of radionuclides since it is based solely on the registration of photons from electron-positron annihilation, but the positronium lifetime imaging ($\tau_{\text{Ps}}$) is possible only when using radionuclides emitting prompt gamma (e.g.\ $^{44}$Sc, $^{52}$Mn, $^{55}$Co, $^{82}$Rb, $^{60}$Cu, $^{124}$I)~\cite{Moskal2020b,Das2023,Rathod2025}. The registration of prompt gamma emitted by radionuclide is needed to reconstruct the time when positron is injected into the tissue, while annihilation photons are used to determine the position and time of the positron annihilation.

The feasibility studies of positronium lifetime imaging were first explored for ortho-positronium annihilations into three photons~\cite{Moskal2019b}, showing that this could be possible with the total-body PET scanners, however, with rather low statistics. A breakthrough on the path to positronium lifetime imaging was the realization that such imaging can be also performed using annihilation of ortho-positronium into two photons \textit{via} pick-off or conversion processes that occurs in the tissue~\cite{Moskal2020a}. Feasibility studies of 2$\gamma$ positronium imaging have shown that it will be over 300 times more effective compared to 3$\gamma$~\cite{Moskal2020a}. This is because in the tissue (i) the rate of ortho-positronium annihilations into 2$\gamma$ is about 70 times higher with respect to 3$\gamma$, (ii) attenuation in the body of 2$\gamma$ is several times lower compared to 3$\gamma$, and (iii) the detection and selection efficiency for 3$\gamma$ is lower than for 2$\gamma$.

In general, simulation-based feasibility studies have demonstrated that positronium imaging is possible and can be performed with upgraded PET scanners after adapting their electronics and data acquisition systems to simultaneous registration of at least two annihilation photons and the prompt gamma~\cite{Moskal2019b,Moskal2020a}. Such a demonstrator of a multi-photon PET scanner, with the triggerless signal acquisition system, was developed and constructed by the J-PET group at the Jagiellonian University~\cite{Moskal2011,Moskal2014,Korcyl2014,Niedzwiecki2017}.

Using the J-PET scanner demonstrator, in 2021 the first experimental 3$\gamma$ images were published with the extended phantoms made from porous materials~\cite{Moskal2021c} demonstrating the possibility of imaging the 3$\gamma$ to 2$\gamma$ annihilation rate ($R_{3/2}$) as a possible diagnostic indicator~\cite{Jasinska2017a}. The same year also the first \textit{ex vivo} positronium lifetime images were achieved for the cardiac myxoma and adipose tissues placed inside the scanner~\cite{Moskal2021d}. These studies demonstrated that positronium lifetime image can be determined simultaneously to the standard PET image~\cite{Moskal2021d} and paved the way for further development of this method. Notably, the obtained \textit{ex vivo} positronium lifetime images served not only as a proof-of-concept of the positronium lifetime imaging method but also revealed a significant difference between the mean ortho-positronium lifetime in cardiac myxoma and healthy adipose tissues~\cite{Moskal2021d,Moskal2023} encouraging further development.

The first \textit{ex vivo} positronium lifetime images were reconstructed directly by identifying annihilation sites, grouping them into voxels and establishing the lifetime spectrum for each voxel separately~\cite{Moskal2021d}. The spatial resolution of such direct imaging is limited by the time-of-flight (TOF) resolution of the scanner. It was shown that a time resolution of 50 ps (coincidence resolving time -- CRT) is required to obtain the point spread function (PSF) of a directly reconstructed 5~mm positronium image~\cite{Moskal2020a}, while best current clinical PET systems achieve 210~ps~\cite{Sluis2019}. This limitation was overcome by developing the first iterative positron lifetime imaging algorithms~\cite{Qi2022}, enabling positronium lifetime imaging with a spatial resolution of 4~mm, even for PET systems with TOF resolution of about 500~ps, as in the case of uEXPLORER total-body PET~\cite{Spencer2021}.

In 2024 the first clinical positronium lifetime images of the human brain were reported~\cite{Moskal2024a}, thus achieving the main milestone on the way of introducing the positronium lifetime imaging into clinics. The first positronium imaging of human was performed at the Warsaw Medical University in Poland using a portable second-generation J-PET scanner~\cite{Moskal2020b,Tayefi2023,Tayefi2024}.

\begin{figure}[htbp]
	\centering
	\includegraphics[width=\linewidth]{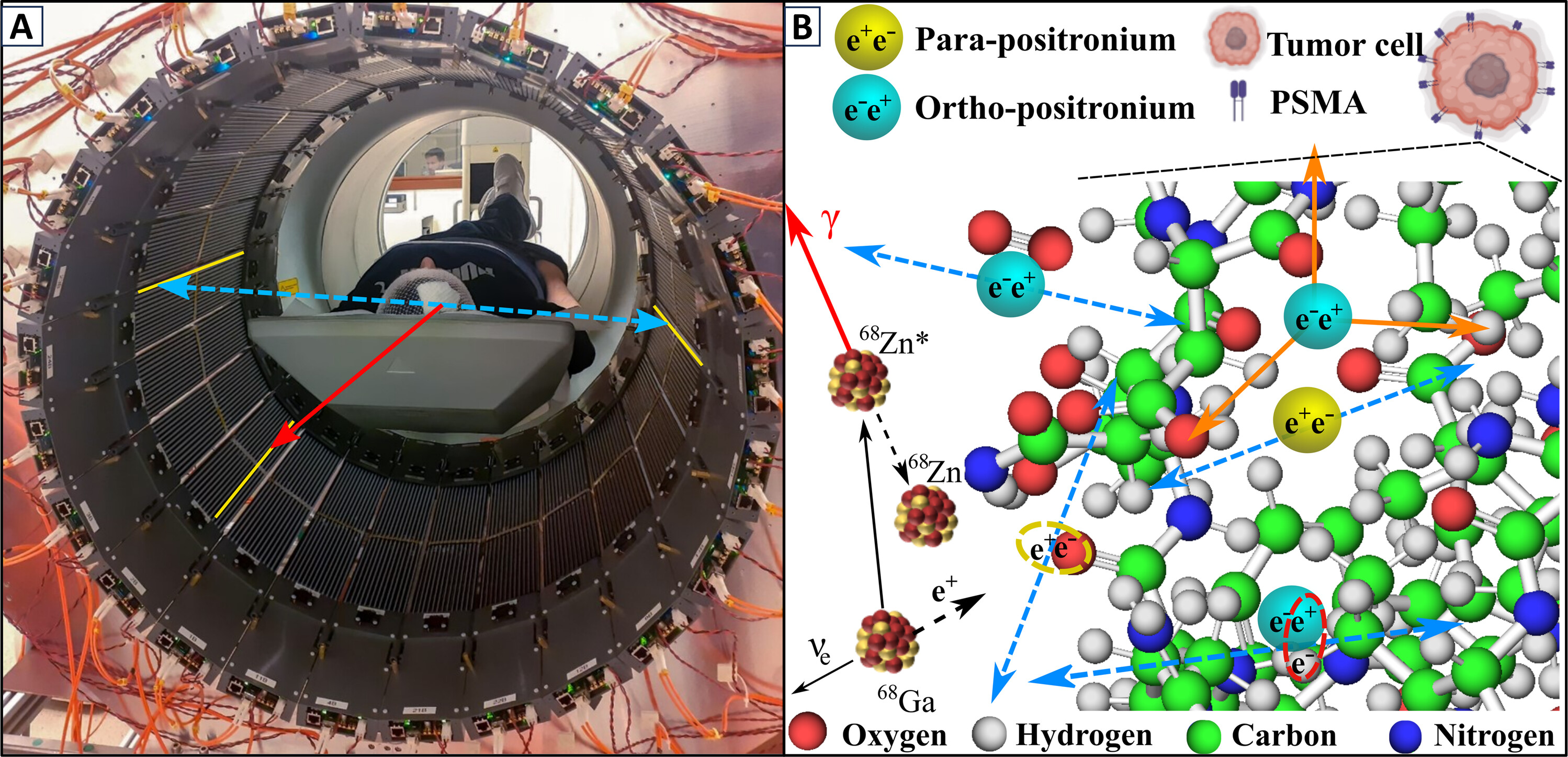}
	\caption{
		(A) Photograph of the patient in the modular J-PET scanner. Blue-dashed arrows indicate photons from electron-positron annihilation, and red-solid arrow indicates prompt gamma from the deexcitation of the $^{68}$Zn* radionuclide. In yellow, those plastic strips of the J-PET tomograph are marked in which the gamma rays interacted.
		(B) Illustration of the $^{68}$Ga radionuclide decay ($^{68}$Ga $\rightarrow$ $^{68}$Zn* + $e^+$ + $\nu$ $\rightarrow$ $^{68}$Zn + $\gamma$ + $e^+$ + $\nu$), followed by positron annihilation within a PSMA (prostate specific membrane antigen) molecule overexpressed by microvascular endothelial cells in glioblastoma multiforme~\cite{Wernicke2011}. Positron-electron annihilation may proceed directly (yellow ellipse), \textit{via} formation of para-positronium (yellow circle), or \textit{via} formation of ortho-positronium (blue circles). Ortho-positronium may self-annihilate into three photons (orange arrows), while para-positronium self-annihilates into two photons (blue-dashed arrows). Ortho-positronium annihilation may also proceed \textit{via} conversion to para-positronium on an oxygen molecule (oPs + O$_2$ $\rightarrow$ pPs + O$_2$ $\rightarrow$ 2$\gamma$ + O$_2$), or \textit{via} pick-off interaction (red ellipse). Blue-dashed arrows indicate two-photon annihilations used for positronium lifetime imaging.
		\textcopyright~2024 American Association for the Advancement of Science. Reprinted with permission from Moskal~\textit{et~al.}~\cite{Moskal2024a} under a Creative Commons Attribution License 4.0 (CC BY).
	}
	\label{fig:jpet_decay}
\end{figure}

These studies were conducted with the $^{68}$Ga-labeled pharmaceuticals intravenously and intratumorally administered to the patient diagnosed with primary brain glioma~\cite{Moskal2024a}. Fig.~\ref{fig:jpet_decay}a shows a photograph of the patient in the J-PET tomograph taken during diagnosis. The superimposed arrows indicate the annihilation photons (dashed) and prompt gamma (solid). This case study not only showed for the first time a clinical positronium lifetime image of human, but importantly, it revealed \textit{in vivo} the differences of the mean lifetime of ortho-positronium ($\tau_{\text{Ps}}$), and mean lifetime of positron ($\tau_{e^+}$) in glioma cancer and healthy brain tissues. The determined positronium $\tau_{\text{oPs}}$, and positron mean annihilation lifetime $\tau_{e^+}$ are smaller in the glioma tumor compared to the healthy brain tissues.

Currently, there are already two clinical long axial field-of-view PET scanners, Biograph Vision Quadra in Bern, Switzerland~\cite{Prenosil2022}, and PennPET Explorer in Philadelphia, USA~\cite{Karp2020,Dai2023}, capable of triggerless (singles mode) data acquisition and hence simultaneous detection of annihilation and prompt photons, and one clinical brain-dedicated TOF-PET scanner VRAIN in Chiba, Japan capable of measuring the positronium lifetime~\cite{Takyu2024a,Takyu2024b}.

In the first step on the way to adopt clinical PET systems to positronium imaging, the Biograph Vision Quadra was used to validate the positronium lifetime measurements~\cite{Steinberger2024} with the certified quartz-glass samples~\cite{Takyu2022} and clinically available radionuclides $^{82}$Rb, $^{68}$Ga, and $^{124}$I~\cite{Steinberger2024}. These measurements were reported soon after publication of the first feasibility studies of $^{124}$I application for positronium lifetime measurement in aqueous solution~\cite{Takyu2023}.

In the next step in 2024, the first whole-body triple coincidence image was demonstrated using Biograph Vision Quadra and ortho-positronium mean lifetime was measured in two patients administered $^{68}$Ga-labelled pharmaceuticals, and one human subject administered $^{82}$Rb-Chloride~\cite{Mercolli2024}. The triple coincidence included events with signals from two 511~keV annihilation photons and one signal from prompt gamma emitted by the radionuclide. $^{82}$Rb-Chloride, as the Food and Drug Administration-approved radiopharmaceutical, was recognized as applicable for the positronium imaging of cardiovascular system~\cite{Moskal2020b,Moskal2023}. It is important to note that in a subject administered [$^{82}$Rb]Cl, the measured mean ortho-positronium lifetime in the right heart ventricle was higher than in the left heart ventricle~\cite{Mercolli2024}, indicating a possible shortening of the lifetime in more oxygenated blood in line with the indications that positronium may be a hallmark of oxygen concentration~\cite{Moskal2021b,Shibuya2020,Stepanov2020}. Notably, the positronium lifetime as a function of oxygen partial pressure in the range between hypoxic and normoxic conditions and quantification of radicals in aqueous solution by positronium lifetime were measured using clinical brain-dedicated TOF-PET scanner VRAIN~\cite{Takyu2024a,Takyu2024b}. Thus, making a step forward towards the clinical use of positronium as a biomarker of hypoxia, a possibility recently indicated in Ref.~\cite{Moskal2021b,Shibuya2020,Stepanov2020}.

Recently also high-resolution experimental positronium lifetime imaging of an extended cylindrical phantom with the PennPET Explorer and with application of iterative time-thresholding reconstruction were reported, showing capability of resolving positronium lifetime in the 6~mm thick polycarbonate insert in the cylindrical phantom~\cite{Huang2024a}.

The first successful demonstration of \textit{in vivo} positronium images of humans with the J-PET scanner~\cite{Moskal2024a}, as well as positronium lifetime measurements~\cite{Steinberger2024,Mercolli2024,Takyu2024a,Takyu2024b} and images~\cite{Huang2024a} demonstrated with clinical PET systems, paves the way for applications of positronium imaging in clinics. However, positronium lifetime imaging is still in its early stages of translation into clinical practice and possesses significant potential for further development. Various research aspects that can lead to its improvement can be grouped into the following categories: (i) performing bio-medical research to enable diagnostic interpretation of the lifetime of positronium atoms in tissues and testing positronium imaging in clinics as a possible biomarker of tissue pathology and hypoxia, (ii) determining the best suited radionuclides that emit positrons and de-excitation gamma quanta ($\beta^+ \gamma$ radionuclides) and developing methods for their effective production, (iii) selecting pharmaceuticals most suitable for positronium imaging and developing methods of attaching $\beta^+ \gamma$ radionuclides to them, (iv) developing iterative algorithms for reconstructing positronium lifetime images, which will enable obtaining high-precision temporal and spatial resolutions, (v) developing methods for effective calibration of scanners in terms of positronium lifetime, and (vi) modernizing PET scanners with the ability to simultaneously record annihilation and deexcitation photons.

In this review we will focus on describing the development status and prospects of positronium lifetime imaging in clinics. We will describe the method of positronium lifetime imaging, present the first \textit{in vivo} positronium lifetime images of the human as well as first measurements of positronium lifetime in humans performed with clinical PET systems. We will describe newly invented method of iterative positronium lifetime imaging and discuss the status and challenges in the above-mentioned research aspects requiring further development.

Complementary recent reviews regarding positronium applications in biology and medicine are given in references~\cite{Bass2023,Moskal2020b,Moskal2022,Hourlier2024}, while review on three gamma imaging in nuclear medicine is given in the reference~\cite{Tashima2024}.

\section{Positronium as a diagnostic parameter}

Standard PET scanners, designed to register electron-positron annihilation into two photons, provide the density distribution of annihilation places in the body (an annihilation intensity image), that in turn informs about the uptake of administered radiopharmaceutical in the body or about the expression on cells of receptors with affinity for the administered radiopharmaceutical~\cite{Alavi2021}. An example of most often used pharmaceuticals for the metabolic imaging is a molecule of fluoro-deoxy-glucose (FDG) with attached positron-emitting radionuclide $^{18}$F enabling imaging of the rate of glycolysis. While the prostate specific membrane antigen (PSMA) labeled with $^{68}$Ga is an example of molecule for imaging of PSMA receptor density at cell membranes.

The mean positronium lifetime delivers information about the tissue molecular nanostructure and concentration of bio-active molecules. This information is different and complementary to the uptake value and receptor density attainable with current PET systems, and also different and complementary to the information about anatomy and morphology attainable with Computed Tomography and Magnetic Resonance, respectively. The following paragraph is meant to explain the physical meaning of the tissue properties that can be sensed by measuring the mean lifetime of positronium.

\begin{figure}[htbp]
	\centering
	\includegraphics[width=\linewidth]{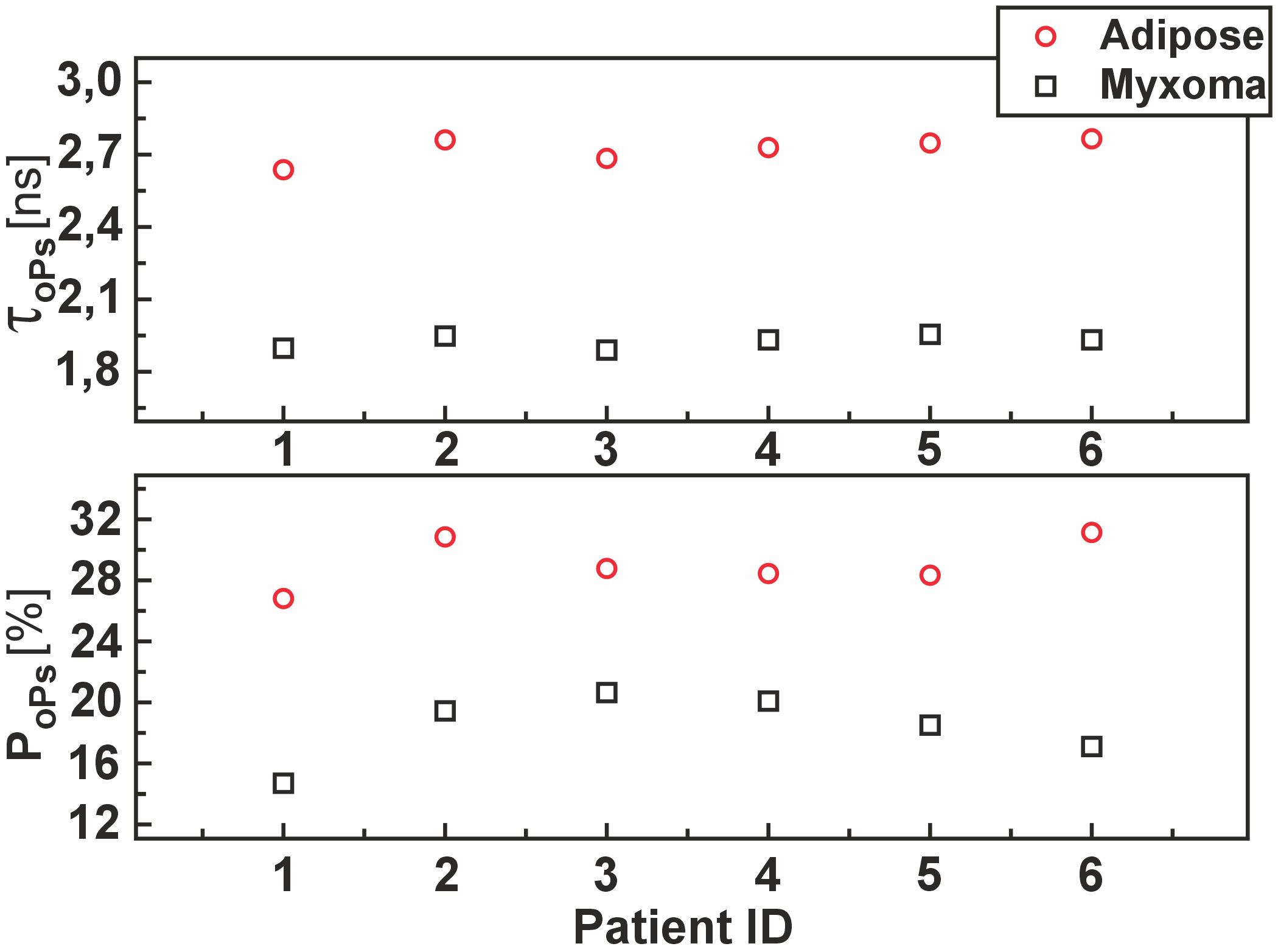}
	\caption{
		Mean lifetime ($\tau_{\mathrm{oPs}}$) and formation probability ($P_{\mathrm{oPs}}$) of ortho-positronium in cardiac myxoma tumor (black squares) and mediastinal adipose tissues (red circle) operated from six patients (ID)~\cite{Moskal2023}. The measurements were performed \textit{ex vivo} using positron annihilation lifetime spectroscopy~\cite{Kubat2024}.
		\textcopyright~2023 Springer Nature. Reprinted with permission from Moskal~\textit{et~al.}~\cite{Moskal2023} under a Creative Commons Attribution 4.0 International License.
	}
	\label{fig:oPs_lifetime}
\end{figure}

Positronium may be formed as the spin-one state referred to as ortho-positronium, or as the spin-zero state referred to as para-positronium. In vacuum, due to the conservation of charge conjugation symmetry, the mean lifetime of para-positronium (125~ps) is more than thousand times shorter than the mean lifetime of ortho-positronium (142~ns)~\cite{Bass2023}. In the tissue, the mean para-positronium lifetime is influenced only by several picoseconds, while mean lifetime of ortho-positronium is shortened substantially, from 142~ns in vacuum to few nanoseconds in the tissue~\cite{Bass2023}. The significant shortening of the ortho-positronium lifetime in the tissue with respect to its value in vacuum is due to the additional possibilities of annihilations, such as annihilation of positron from positronium \textit{via} interaction with electrons (pick-off process), or the spin exchange with the paramagnetic molecules such as oxygen (conversion process). These processes are indicated pictorially in Fig.~\ref{fig:jpet_decay}b. The mean lifetime of ortho-positronium varies between 1.4~ns and 2.9~ns depending on the tissue type~\cite{Ahn2021, Moskal2023, Jean2006, Chen2012, Avachat2024, Jasinska2017b, Jasinska2017c, Zgardzinska2020, Moyo2022, Karimi2023, Moskal2021d}. Recently reported differences in $\tau_{\text{oPs}}$ between myxoma tumor and adipose tissues operated from the patients are at the level of 800~ps as shown in Fig.~\ref{fig:oPs_lifetime}~\cite{Moskal2023, Moskal2021d}. Fig.~\ref{fig:oPs_lifetime} shows also that the ortho-positronium production probability $P_{\text{oPs}}$ substantially differs between myxoma tumor and adipose tissues.

In general, the mean ortho-positronium lifetime provides information about the size of free volumes between atoms and the concentration of the bioactive molecules such as oxygen molecules, while the positronium production probability informs about the concentration of free spaces between atoms (voids). The medical meaning of these parameters and their correlation with the kind of the tissue pathology remains to be explored.

For the diagnostic purposes important are changes in the values of parameters $\tau_{\text{oPs}}$, $\tau_{e^+}$, $\tau_d$, $R_{3/2}$, and $P$, as a function of the degree of the tissue pathology for the same tissue type. The \textit{in vitro} studies of healthy and cancer cells, cell spheroids and tissues~\cite{Karimi2023, Axpe2014, Jasinska2017b}, show such differences in the mean ortho-positronium lifetime between healthy and cancerous cells and tissues of the same type but of different degree of malignancy, at the level of 50~ps. Thus, indicating that the concentration and size of the intramolecular and intermolecular voids sensed by positronium may be a hallmark of the early molecular alteration of the tissue.

\begin{figure}[htbp]
	\centering
	\includegraphics[width=\linewidth]{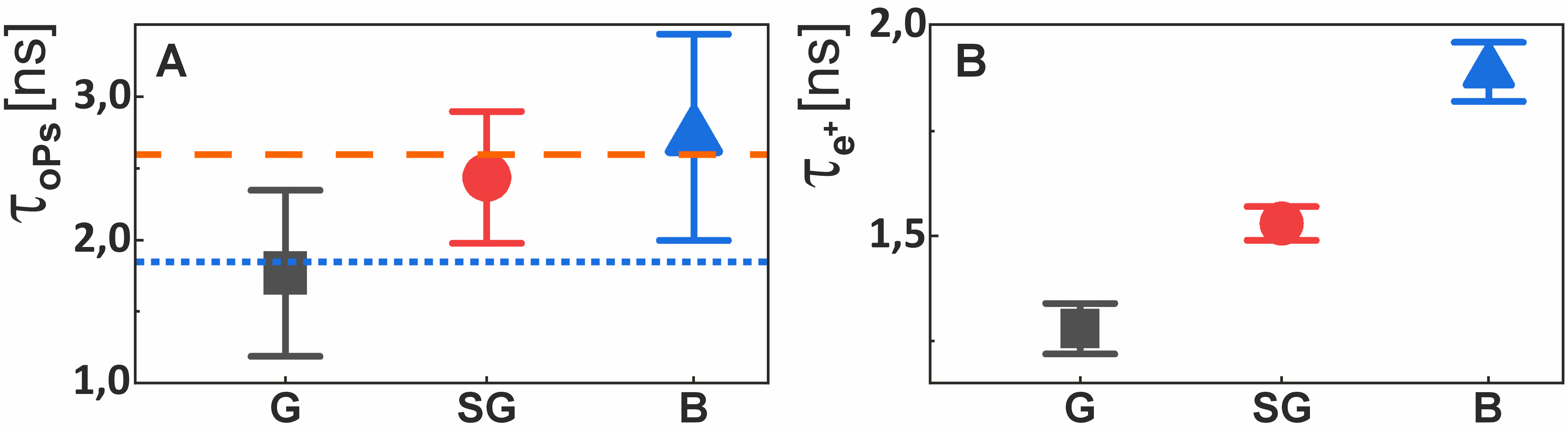}
	\caption{
		Mean ortho-positronium lifetime, $\tau_{\mathrm{oPs}}$, (A), and mean positron lifetime, $\tau_{e^+}$, in the range of 0 to 5~ns (B), determined \textit{in vivo} for a patient with glioma tumor using the modular J-PET scanner~\cite{Moskal2024a}. 
		Black squares represent measurements for glioblastoma tumor (G), red circles for salivary glands (SG), and blue triangles for healthy brain tissues (B). 
		The \textit{in vivo} data are compared with the \textit{ex vivo} results: 1.92~ns for cardiac myxoma tumor (blue-dotted line) and 2.72~ns for adipose tissue (orange-dashed line), as presented in Fig.~\ref{fig:oPs_lifetime}~\cite{Moskal2023}.
		\textcopyright~2024 American Association for the Advancement of Science. 
		Reprinted with permission from Moskal~\textit{et~al.}~\cite{Moskal2024a} under a Creative Commons Attribution License 4.0 (CC BY).
	}
	\label{fig:oPs_in_vivo}
\end{figure}

The \textit{ex vivo} results are promising but the translation of positronium as a diagnostic biomarker to clinics requires studies in the \textit{in vivo} conditions. The first \textit{in vivo} images of $\tau_{\text{oPs}}$ in human body~\cite{Moskal2024a} shown in Fig.~\ref{fig:oPs_in_vivo}a indicate the same tendency as observed \textit{ex vivo}, showing differences of $\tau_{\text{oPs}}$ between healthy brain and glioma tumor tissues, though this first result is based on the very low statistics. Notably, even in this first clinical imaging the statistically significant differences between the glioma and healthy brain are observed for the mean lifetime of positron, $\tau_{e^+}$, as it is shown in Fig.~\ref{fig:oPs_in_vivo}b.

The mean positron lifetime in the tissue, $\tau_{e^+}$, combines information from all possible annihilation mechanisms. It may be seen as a counts-weighted average of para-positronium ($\tau_{\text{pPs}}$), ortho-positronium ($\tau_{\text{oPs}}$) and direct annihilations ($\tau_d$) lifetimes. Therefore, $\tau_{e^+}$ is an effective parameter depending on the intramolecular void size and oxygen concentration (primarily influencing $\tau_{\text{pPs}}$) and bulk molecular structure (primarily influencing $\tau_d$)~\cite{Bass2023}.

\begin{figure}[htbp]
	\centering
	\includegraphics[width=\linewidth]{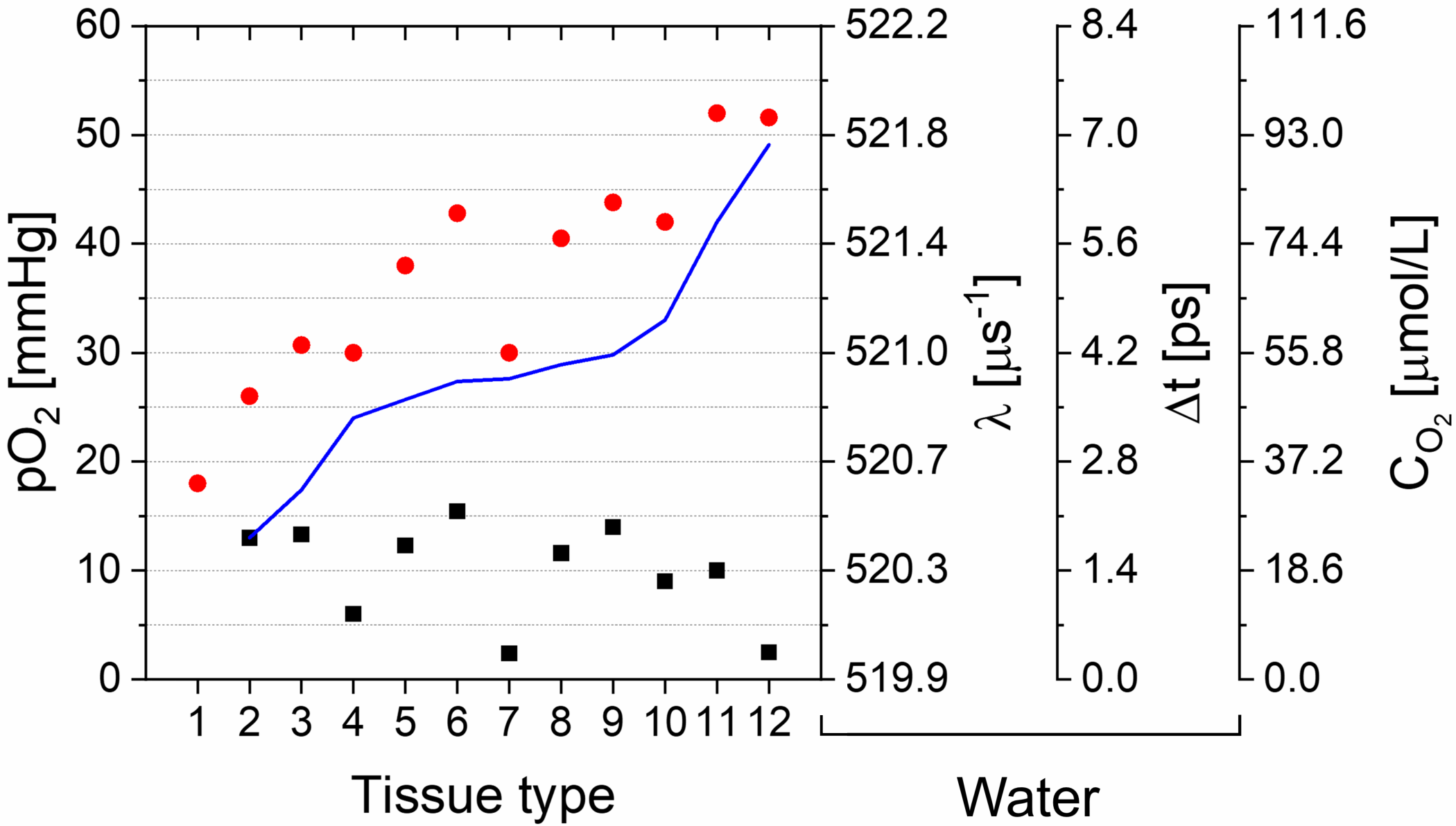}
	\caption{
		The mean value of partial pressure of oxygen molecules (pO$_2$) is shown for tissues of: 
		1 -- myocardium, 
		2 -- brain, 
		3 -- kidney, 
		4 -- liver, 
		5 -- head and neck, 
		6 -- lung, 
		7 -- prostate, 
		8 -- skin melanoma, 
		9 -- sarcoma, 
		10 -- cervical, 
		11 -- breast, 
		12 -- pancreas. 
		The shown data refer to medians compiled in~\cite{Moskal2021b}. 
		Red circles indicate values for healthy tissues, while black squares indicate values for cancer tissues. 
		Tissues are ordered according to the increasing difference in oxygen partial pressure, shown as a blue curve.
		Axes on the right indicate: the ortho-positronium annihilation rate in water ($\lambda$), the difference ($\Delta t$) between ortho-positronium mean lifetime at a given pO$_2$ value and at pO$_2 = 0$, and the concentration of oxygen in water (C$_{O_2}$).
		\textcopyright~2024 Index Copernicus Ltd.
		Reprinted with permission from Moskal~\textit{et~al.}~\cite{Moskal2021b} under a Creative Commons Attribution License 4.0 (CC BY).
	}
	\label{fig:pO2_tissues}
\end{figure}

Recently, also the studies of the ortho-positronium mean lifetime dependence on the oxygen concentration in water and some organic liquids were reported~\cite{Stepanov2020, Shibuya2020, Takyu2024a}. The differences between normoxic and hypoxic oxygen pressure in tissues are summarized in Ref.~\cite{Moskal2021b} and shown in Fig.~\ref{fig:pO2_tissues}. The figure indicates that on the average the partial oxygen pressure in cancer tissues is smaller by about 10~mmHg to 50~mmHg with respect to the healthy tissues of the same kind. It is presently not known how much the ortho-positronium lifetime in the living tissue is increased when the oxygen partial pressure decreases by 50~mmHg. However, taking into account the linear dependence of the ortho-positronium decay rate on the partial pressure of oxygen in the water and some organic liquids~\cite{Shibuya2020, Stepanov2020}, it can be estimated that in water, as it is indicated in Fig.~\ref{fig:pO2_tissues}, the changes would be rather small, in the order of 5~ps only. In organic liquids, however, the change of oxygen pressure by 50~mmHg causes decrease of ortho-positronium lifetime by 250~ps in isopropanol organic liquid and as much as 540~ps in isooctane~\cite{Stepanov2020, Shibuya2020}. Isopropanol and isooctane are examples of organic molecules similar in structure to most of typical cell metabolites.

The measurement of hypoxia with positronium will be then challenging by the heterogeneity of the tissue molecular structure that causes the changes in the order of e.g.\ 50~ps in cardiac myxoma tumor and 100~ps in healthy adipose tissues~\cite{Moskal2023, Moskal2021d} as can be seen in Fig.~\ref{fig:oPs_lifetime}.

As regards the \textit{in vivo} assessment of oxygen concentration, very promising results were obtained from the first \textit{in vivo} measurements of the ortho-positronium lifetime by Biograph Vision Quadra of the subject administered [$^{82}$Rb]Cl~\cite{Mercolli2024}. The lifetime of ortho-positronium in the right heart ventricle was observed to be on the average 1.96~ns, compared to 1.44~ns in the left ventricle. The difference may be attributed to the higher oxygenation level of blood in left heart ventricle compared to the right one. Due to the ortho-positronium to para-positronium conversion process (oPs + O$_2$ $\rightarrow$ pPs + O$_2$ $\rightarrow$ 2$\gamma$ + O$_2$), with the increasing oxygen concentration, the ortho-positronium lifetime is decreasing.

\section{A method of positronium lifetime imaging}

Positronium imaging~\cite{Moskal2019a} is a novel technique in positron emission tomography (PET) that leverages the unique properties of positronium, a bound state of an electron and a positron, to gather more detailed information about the molecular environment in biological tissues. This method aims at the early identification of disease progression by analyzing the lifetimes and annihilation characteristics (such as $\tau_{\text{oPs}}$, $\tau_{e^+}$, $\tau_d$, $R_{3/2}$, and $P$) of positronium atoms formed during PET scanning. The technique was initially conceptualized~\cite{Moskal2019a}, validated with simulations~\cite{Moskal2019b, Moskal2020a}, and demonstrated experimentally using the J-PET (Jagiellonian Positron Emission Tomograph) system both for positron and positronium lifetime~\cite{Moskal2021d, Moskal2024a} and for the images of 3$\gamma$ annihilations for $R_{3/2}$ parameter~\cite{Moskal2021c}. 

Positronium lifetime imaging is a subclass of the positronium imaging combining the Positron Emission Tomography (PET) and Positronium Annihilation Lifetime Spectroscopy (PALS)~\cite{Moskal2019a}. PALS is a spectroscopic method commonly applied in material science~\cite{Bass2023, Gidley2006, Zahasky2018}, enabling to study porosity of material samples at the nanometer and sub-nanometer scale by the measurement of the positron lifetime spectrum in the studied material. PET is a diagnostic method to determine the density distribution of positron annihilations in the patient body. The positronium lifetime imaging method delivers spatially resolved information (images) of the mean positron lifetime ($\tau_{e^+}$), positronium lifetime ($\tau_{\text{oPs}}$), mean lifetime of positron before direct annihilation ($\tau_d$), as well as images of the probability of positronium formation ($P$) in the examined object.

A key requirement for positronium lifetime imaging is the detection of both annihilation photons and prompt (de-excitation) photon from the decay of radionuclide. This necessitates the use of multi-photon tomography systems like J-PET capable of registering more than two photons in coincidence~\cite{Moskal2024a}, and application of radionuclides like $^{22}$Na which emit prompt gamma rays ($\gamma$) alongside positrons ($e^+$) 
(${}^{22}\text{Na} \rightarrow {}^{22}\text{Ne}^* + e^+ + \nu \rightarrow {}^{22}\text{Ne} + \gamma + e^+ + \nu$),
where $\gamma$ serves as a time marker for the annihilation process. The $^{22}$Na radionuclide is commonly used in PALS studies; however, its half-life of 2.6 years (Table~\ref{tab:isotopes_properties}) makes it impractical for applications in medical diagnostics. Therefore, other radionuclides such as $^{44}$Sc, $^{52}$Mn, $^{55}$Co, $^{68}$Ga, $^{82}$Rb, $^{124}$I are explored in view of positronium imaging due to their already established or potential clinical relevance~\cite{Moskal2020a}.

\begin{table*}[t]
	\centering
	\caption{Physical properties of selected $\beta^+\gamma$ radionuclides considered for positronium imaging. The table incorporates information on the half-life ($t_{1/2}$), the branching ratio of the positrons ($Y_{\beta^+}$), the maximum positron energy ($E_{\text{max}}$), the mean range ($R_{\text{mean}}$) and maximum range ($R_{\text{max}}$) of positron in water~\cite{Mryka2023}, the calculated value of the mean range are consistent within 10\% with the results obtained in reference~\cite{Kertesz2022}, the energy of the prompt gamma ($E_{\gamma}$), the branching ratio of the prompt gamma ($Y_{\gamma}$), the branching ratio of the pure $\beta^+ + \gamma$ events removing the electron capture contribution ($Y_{\beta^+ + \gamma}$) (subject to large errors)~\cite{IAEA}, the median time interval between the emission of positron and prompt gamma ($t_{1/2}^{\text{deexc}}$) and the maximum energy of electron achievable in the Compton scattering of the prompt gamma with the electron (Comp. E.) for selected isotopes for positronium technique~\cite{Das2023,Sitarz2020,Conti2016,George2021,Matulewicz2021,Dong2015}.}
	\label{tab:isotopes_properties}
	\renewcommand{\arraystretch}{1.2}
	\setlength{\tabcolsep}{4pt}
	\begin{tabular}{|l|c|c|c|c|c|c|c|c|c|c|c|}
		\hline
		\textbf{Isotope} & \textbf{$t_{1/2}$} & \textbf{$Y_{\beta^+}$} & \textbf{$E_{\text{max}}$} & \textbf{$R_{\text{mean}}$} & \textbf{$R_{\text{max}}$} & \textbf{$E_\gamma$} & \textbf{$Y_\gamma$} & \textbf{$Y_{\beta^+ + \gamma}$} & \textbf{$Y_{\beta^+ + \gamma}/Y_{\beta^+}$} & \textbf{$t^{\text{deexc}}_{1/2}$} & \textbf{Comp. E.} \\
		&  & \textbf{[\%]} & \textbf{[MeV]} & \textbf{[mm]} & \textbf{[mm]} & \textbf{[MeV]} & \textbf{[\%]} & \textbf{[\%]} & \textbf{[\%]} & \textbf{[ps]} & \textbf{[MeV]} \\
		\hline
		$^{22}$Na  & 2.60 y  & 89.95 & 0.545 & 0.54 & 1.85 & 1.275 & 99.94 & 89.90 & 99.94 & 3.6   & 1.062 \\
		$^{52}$Mn  & 5.6 d   & 29.40 & 0.575 & 0.58 & 1.99 & 1.434 & 100   & 30    & 100   & 0.78  & 1.217 \\
		&         &       &       &      &      & 0.936 & 94.5  & 29    & 98.69 & 6.7   & 0.735 \\
		&         &       &       &      &      & 0.744 & 90    & 29    & 98.69 & 41    & 0.553 \\
		$^{124}$I  & 4.17 d  & 22.69 & 1.822 & 2.15 & 8.53 & 0.603 & 62.9  & 12    & 52.89 & 6.2   & 0.423 \\
		&         &       &       &      &      & 0.723 & 10.36 & 0.25  & 4.58  & 1.04  & 0.534 \\
		$^{72}$As  & 26.0 h  & 87.86 & 3.33  & 4.27 & 16.64& 0.834 & 81    & 71    & 80.81 & 3.35  & 0.638 \\
		$^{55}$Co  & 17.53 h & 75.89 & 1.5   & 1.72 & 6.79 & 0.931 & 75    & 59    & 77.74 & 8     & 0.731 \\
		&         &       &       &      &      & 0.477 & 20.2  & 13.5  & 17.79 & 37.9  & 0.311 \\
		&         &       &       &      &      & 1.409 & 16.9  & 11.3  & 14.89 & 37.9  & 1.192 \\
		$^{44}$Sc  & 3.97 h  & 94.3  & 1.474 & 1.69 & 6.65 & 1.157 & 99.9  & 94.3  & 100   & 2.61  & 0.948 \\
		$^{68}$Ga  & 67.71 m & 88.91 & 1.899 & 2.25 & 8.94 & 1.077 & 3.22  & 1.19  & 1.34  & 1.57  & 0.870 \\
		$^{60}$Cu  & 23.7 m  & 92.59 & 0.653 & 0.67 & 2.37 & 1.333 & 88    & 81    & 87.48 & 0.735 & 1.118 \\
		&         &       &       &      &      & 1.791 & 45.4  & 42    & 45.36 & 0.24  & 1.567 \\
		&         &       &       &      &      & 0.826 & 21.7  & 20.4  & 22.03 & 0.59  & 0.631 \\
		$^{82}$Rb  & 1.26 m  & 95.36 & 3.382 & 4.34 & 16.92& 0.777 & 15.1  & 13.5  & 14.16 & 4.45  & 0.584 \\
		$^{14}$O   & 70.6 s  & 99.89 & 1.808 & 2.13 & 8.45 & 2.313 & 99.39 & 99.26 & 99.38 & 0.068 & 2.083 \\
		$^{10}$C   & 19.3 s  & 99.97 & 2.93  & 3.69 & 14.51& 0.718 & 100   & 99.97 & 100   & 710   & 0.530 \\
		\hline
	\end{tabular}
\end{table*}

\begin{figure}[htbp]
	\centering
	\includegraphics[width=\linewidth]{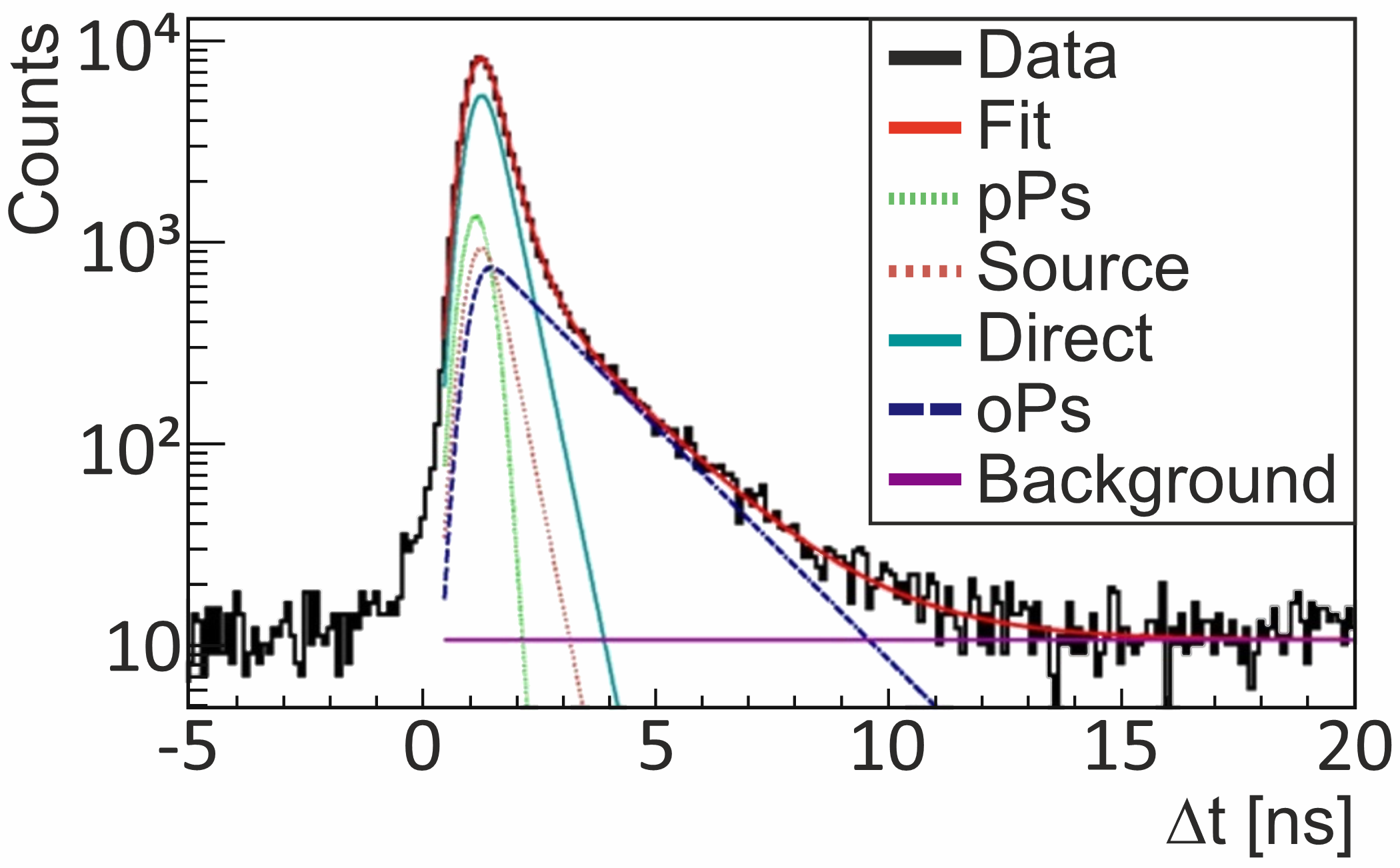}
	\caption{
		Distribution of positron lifetime $\Delta t$ determined in cardiac myxoma tissue.
		Superimposed curves correspond to:
		para-positronium (light green dashed line),
		ortho-positronium (dark blue dashed line),
		direct annihilation of the positron and electron without positronium formation (turquoise line),
		annihilations in the source material (red dashed line),
		and accidental coincidences (violet line).
		Annihilation in the sources is present in the \textit{ex vivo} studies.
		The sum of all contributions is shown as the red solid line.
		The fitted function is a convolution of exponential and Gaussian functions plus a constant background.
		\textcopyright~2021 American Association for the Advancement of Science.
		Reprinted with permission from Moskal~\textit{et~al.}~\cite{Moskal2021d} under a Creative Commons Attribution License 4.0 (CC BY).
	}
	\label{fig:lifetime_distribution}
\end{figure}

In positronium lifetime imaging, the time and place $(t_a, x_a, y_a, z_a)$ of electron-positron annihilation in the body are reconstructed based on the registration of the time and position of interaction in the detector of annihilation photons $(t_1, x_1, y_1, z_1)$ and $(t_2, x_2, y_2, z_2)$. Currently, only electron-positron annihilation into two photons is used (see blue dashed arrows in Fig.~\ref{fig:jpet_decay}). The values of $t_1$ and $t_2$ are measured relative to the internal electronics clocks, so only a time difference $t_2 - t_1$, referred to as time-of-flight (TOF), is of physical meaning. The lifetime of a positron in the body is reconstructed as the time interval $\Delta t$ between the time of prompt gamma emission, $t_p$ (red arrow in Fig.~\ref{fig:jpet_decay}), and the time of annihilation photon creation, $t_a$. 
As a result of the measurement a list-mode data in the form of $(\Delta t, \text{TOF}, x_1, y_1, z_1, x_2, y_2, z_2)$ are created and used for positronium lifetime image reconstruction. In the first positronium imaging, a direct reconstruction method was applied~\cite{Moskal2021d, Moskal2024a}. In this method, for each event the annihilation place and time $(\Delta t, x_a, y_a, z_a)$ are reconstructed and the annihilation density image of $(x_a, y_a, z_a)$ points is created. Next, for each voxel of this image, a $\Delta t$ spectrum is formed, and this spectrum is used to extract the values of parameters such as $\tau_{\text{oPs}}$, $\tau_{e^+}$, $\tau_d$, and $P$. The values of these parameters are extracted for each voxel separately, enabling to determine images such as $\tau_{\text{oPs}}(x_a,y_a,z_a)$, $\tau_{e^+}(x_a,y_a,z_a)$, $\tau_d(x_a,y_a,z_a)$, and $P(x_a,y_a,z_a)$. Fig.~\ref{fig:lifetime_distribution} shows an exemplary spectrum of $\Delta t$ registered in the cardiac myxoma tumor during the first \textit{ex vivo} positronium imaging~\cite{Moskal2021d}. The superimposed curves are results of the fit deconvoluting the contributions from various annihilation mechanisms, using programs as e.g.\ described in Ref.~\cite{Dulski2018}.

\begin{figure}[htbp]
	\centering
	\includegraphics[width=\linewidth]{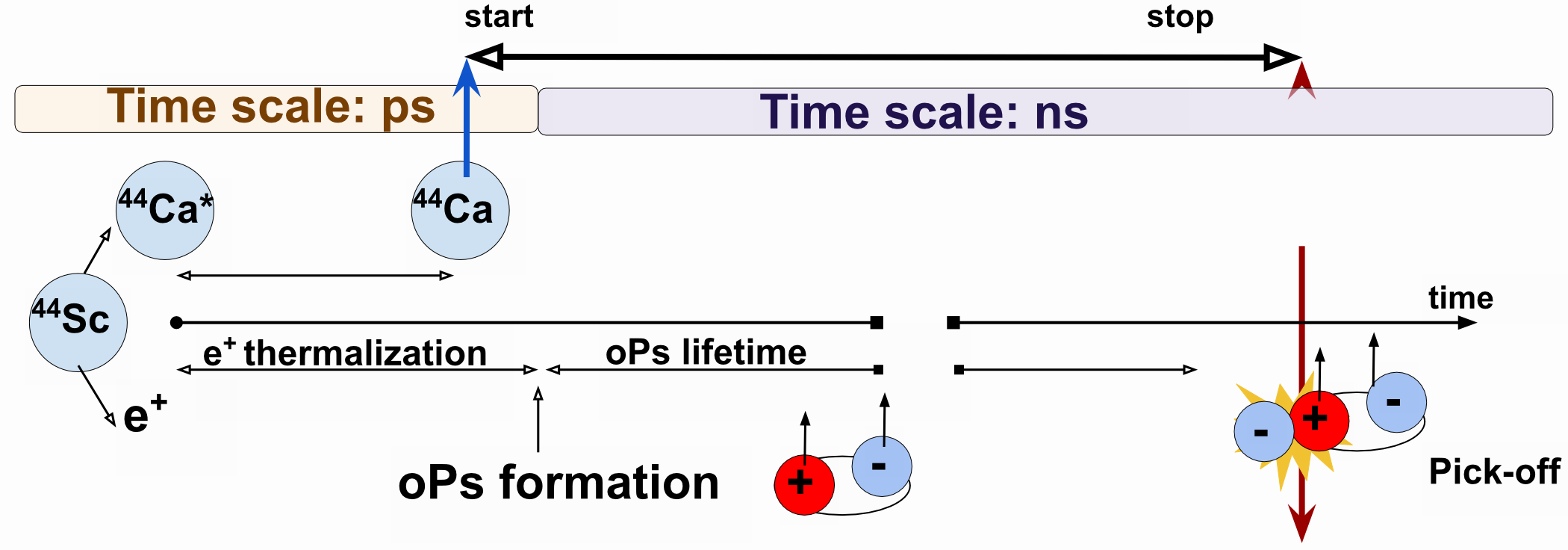}
	\caption{
		Schematic diagram of the temporal sequence of processes in positronium imaging using the example of $^{44}$Sc radionuclide.
		A $^{44}$Sc nucleus undergoes $\beta^+$ decay to an excited $^{44}$Ca* nucleus while emitting a positron. 
		The excited $^{44}$Ca* emits a prompt gamma photon with energy of 1160~keV (dotted blue arrow), on average after about 2.6~ps (see Table~\ref{tab:isotopes_properties}).
		The emitted positron undergoes thermalization in matter, which takes about 10~ps~\cite{Bass2023}, and may subsequently form ortho-positronium.
		Interaction of ortho-positronium with surrounding molecules \textit{via} pick-off process or conversion into para-positronium leads to emission of two photons.
		The mean lifetime of ortho-positronium in tissue is on the order of a few nanoseconds.
		\textcopyright~2020 Springer Nature.
		Reprinted with permission from Moskal~\textit{et~al.}~\cite{Moskal2020a} under a Creative Commons Attribution 4.0 International License.
	}
	\label{fig:temporal_sequence}
\end{figure}

Radionuclides applicable for positronium lifetime imaging, as in standard PET, must possess physical and chemical properties enabling to use them in standard positron emission tomography. Thus, e.g., the half-lifetime must be long enough for conveniently attaching them to pharmaceuticals and performing imaging, but it has to be not too long so the injected radionuclide can be fully cleaned from the patient body within few days. However, in addition they have to emit prompt gamma, and importantly the deexcitation of the daughter nucleus, has to be in a short time, at best in the range of few picoseconds to be substantially less than the time resolution of $\Delta t$ measurement. Otherwise, if the deexcitation time was larger than the time resolution, the unknown deexcitation time would worsen the resolution of determining the lifetime parameters $\tau_{\text{oPs}}$, $\tau_{e^+}$, and $\tau_d$, since the assumption that the emission time of prompt gamma is equal to the emission time of positron would not work.

Each of the above-mentioned isotopes ($^{44}$Sc, $^{52}$Mn, $^{55}$Co, $^{68}$Ga, $^{82}$Rb, $^{124}$I) and others included in Table~\ref{tab:isotopes_properties} present specific challenges that limit their effectiveness. The physical properties important for positronium lifetime imaging were discussed in detail in Ref.~\cite{Das2023} and are listed in Table~\ref{tab:isotopes_properties}. A good isotope for positronium imaging should be characterized by
(i) as short as possible deexcitation time -- $t^{1/2}_{\text{deex}}$ in Table~\ref{tab:isotopes_properties}, 
(ii) large fraction of decays where positron is accompanied by the emission of prompt gamma denoted in Table~\ref{tab:isotopes_properties} as $Y_{\beta^+ + \gamma} / Y_{\beta^+}$ (at best each positron emission should be followed by the prompt gamma emission $Y_{\beta^+ + \gamma} / Y_{\beta^+} = 100\%$),
(iii) energy of the prompt gamma substantially larger than 511~keV to enable efficient discrimination from annihilation photons.

For example, one significant limitation of using $^{68}$Ga is its low branching ratio for positron emission accompanied by prompt gamma-ray emission, which is only $Y_{\beta^+ + \gamma} / Y_{\beta^+} = 1.34\%$. This reduces the probability of detecting the required multi-photon coincidences, leading to weaker signals for positronium lifetime measurements and large background from $2\gamma + \gamma_{\text{scattered}}$ events that may mimic $2\gamma + \gamma_{\text{prompt}}$. Nevertheless, $^{68}$Ga is commonly used in clinics, and the first \textit{in vivo} studies of positronium lifetime imaging were performed with $^{68}$Ga-labeled pharmaceuticals~\cite{Moskal2024a}.

Radionuclide $^{44}$Sc, with the main decay chain
${}^{44}\text{Sc} \rightarrow {}^{44}\text{Ca}^* + e^+ + \nu \rightarrow {}^{44}\text{Ca} + \gamma + e^+ + \nu$,
has emerged as a promising alternative for positronium imaging due to its favorable properties~\cite{Moskal2020b, Moskal2024b}. $^{44}$Sc emits a prompt gamma-ray after each positron emission ($Y_{\beta^+ + \gamma} / Y_{\beta^+} = 100\%$), with a mean deexcitation time of only 2.6~ps (Table~\ref{tab:isotopes_properties}). The relatively long half-life of $^{44}$Sc ($\sim$4~h) also makes it compatible with clinical workflows, allowing for extended imaging protocols without compromising signal strength. Given these advantages, $^{44}$Sc isotope is considered a strong candidate for advancing positronium imaging~\cite{Moskal2020b}, however further development is still needed towards FDA approval of $^{44}$Sc-labeled pharmaceuticals.

\section{Radioisotopes and ligands \\for positronium imaging}

In addition to the physical properties, discussed above, that are required to perform positronium lifetime image reconstruction, the radionuclide must possess specific physical and chemical properties enabling to attach it to the pharmaceutical. In particular, in case of biomolecules, the decay time should be sufficient for diagnostic procedures, meaning that for biomolecules with rapid pharmacokinetics such as peptides and small biologically active molecules, the optimal half-life of $\beta^+$ emitters should range from one to several hours. For the proteins, having slow pharmacokinetics the half-life should be longer than three days. However, the half-life should not be too long to minimize unwanted radiation after the medical procedure.

Recent articles~\cite{Sitarz2020,Das2023} summarize the properties, applications, and production possibilities of radionuclides that can be used in the new coincidence $\beta^+\gamma$ PET technique. These radioisotopes include: $^{10}$C, $^{14}$O, $^{22}$Na, $^{34\text{m}}$Cl, $^{44}$Sc, $^{48}$V, $^{52\text{m}}$Mn, $^{55}$Co, $^{60}$Cu, $^{66}$Ga, $^{68}$Ga, $^{69}$Ge, $^{72}$As, $^{76}$Br, $^{82}$Rb, $^{86}$Y, $^{94\text{m}}$Tc, $^{110\text{m}}$In, and $^{124}$I. The most widely used PET-nuclide for antibody labeling is $^{89}$Zr~\cite{Laforest2016}, which has an appropriate $t_{1/2}=78.42$ h and emits $\beta^+$ radiation with energy $E_{\beta^+\text{max}}=0.902$ MeV, but in only 23\%. Unfortunately, in the decay chain of $^{89}$Zr 
($^{89}$Zr $\rightarrow$ $^{89\text{m}}$Y + $e^+$ + $\nu$ $\rightarrow$ $^{89}$Y + $\gamma$ + $e^+$ + $\nu$)
the mean time between the emission of positron and prompt gamma is about 23 seconds, which makes it impossible to use in the $\beta^+\gamma$ coincidence technique~\cite{Das2023}.

Positronium lifetime imaging was first demonstrated with $^{22}$Na radionuclide~\cite{Moskal2021d} that is clinically not applicable due to its long half-lifetime (2.6 years) and its bone incorporation~\cite{Samuels2021}. However, by now the demonstration of positronium lifetime measurements using clinical isotopes as $^{68}$Ga~\cite{Steinberger2024}, $^{82}$Rb~\cite{Steinberger2024}, and $^{124}$I~\cite{Takyu2023, Steinberger2024} have been already reported for certified reference materials~\cite{Takyu2022}. $^{124}$I radionuclide was also applied for the positronium lifetime images of phantoms~\cite{Mercolli2025a}, while in the first in-human studies $^{68}$Ga was used for imaging~\cite{Moskal2024a}, and $^{82}$Rb and $^{68}$Ga for the first in-human positronium measurement with clinical PET system~\cite{Mercolli2024}.

Here, taking into account criteria related to the pharmacokinetics of the guided vectors and production costs, we limited the detailed description of labelling possibilities to the following three examples: 
$^{44}$Sc for peptides and small biologically active molecules,
$^{55}$Co for larger molecules such as antibody fragments, e.g., affibody and nanobody,
and $^{124}$I for full-length monoclonal antibody. Below, we present their basic properties, production methods, and current application in PET diagnosis.

\subsection*{$^{44}$Sc radionuclide}
Among radioisotopes with a half-life of several hours that can be used to label biomolecules with fast pharmacokinetics, $^{44}$Sc is an ideal candidate for the $\beta^+\gamma$ coincidence. Its half-life is 3.97 h and emits, next to $\beta^+$ (94.27\% abundance), only one $\gamma$ line of high intensity with energy (1157 keV, 99\%). Two production routes can be used to obtain clinically relevant activities. Irradiation of natural or $^{44}$Ca-enriched calcium targets (CaO or CaCO$_3$) with protons in the range of 16$\rightarrow$0 MeV, or use of a $^{44}$Ti/$^{44}$Sc radionuclide generator. By irradiating an enriched $^{44}$CaCO$_3$ target with a proton beam, up to 150 MBq/$\mu$Ah of $^{44}$Sc can be obtained with less than 1\% radioactive impurities~\cite{vanderMeulen2020}. By 2 hours of target irradiation, it is possible to obtain activity sufficient to perform several dozen diagnostic procedures. $^{44}$Sc can be easily isolated from natCa or $^{44}$Ca targets using ion exchange or extraction resins and by Sc(OH)$_3$ precipitation~\cite{Pruszynski2010}. The fastest and easiest to automate are the methods developed using UTEVA~\cite{Valdovinos2015} and DGA~\cite{Mueller2013} extraction resins. The unique properties of $^{44}$Sc make it very suitable for clinical PET applications. They also allow for the transportation of $^{44}$Sc-labeled radiopharmaceuticals to hospitals that are situated several hundred kilometers away from the site where the radiopharmaceuticals are produced.

Another potential source of $^{44}$Sc is the $^{44}$Ti/$^{44\text{g}}$Sc generator, as suggested by Filosofov and Rösch~\cite{Filosofov2010}. The advantage of this solution is the ability to place the generator directly in the hospital~\cite{Gajecki2023}. However, the low reaction cross section of $^{45}$Sc(2,n)$^{44}$Ti and the long half-life of $^{44}$Ti (59 years) mean that in order to obtain 500 MBq of $^{44}$Ti, it is necessary to irradiate the target with high current and energy protons of (560 MBq, 30 MeV, 1 mA, 7 days)~\cite{Roesch2012}. It would be, however important to advanced the technology of $^{44}$Ti/$^{44\text{g}}$Sc preparation not only for positronium imaging, but also for making PET diagnostics more accessible and affordable in low- and medium-income countries by combining the low-costs J-PET technology with the long lived $^{44}$Ti/$^{44}$Sc generator~\cite{Moskal2024b}.

The best chelator for attaching scandium radionuclides to biomolecules is the macrocyclic DOTA ligand (1.2,2',2'',2'''-(1,4,7,10-tetraazacyclododecane-1,4,7,10-tetraethyl)tetraacetic acid) forming a kinetically inert and thermodynamically stable complex with $\log K_{\text{ScDOTA}}=29.0$ which is comparable with stability constants for Y$^{3+}$ and heaviest lanthanides but is higher than those for In$^{3+}$ and Ga$^{3+}$~\cite{Majkowska2011, Pruszynski2012}. Since 2008, when the first papers on the use of $^{44}$Sc in the PET technique began to appear~\cite{Haddad2008}, 156 papers have been published on this topic showing a growing interest in this radionuclide. Given the half-life $^{44}$Sc, which is well matched to the pharmacokinetic profile of peptides, and other small biologically active molecules, $^{44}$Sc is considered a promising radiometal for PET imaging. So far, preclinical studies have been conducted on the following bioconjugates: DOTA-bombesin~\cite{Koumarianou2012}, DOTA octreotide~\cite{Domnanich2017}, DOTA-folic acid~\cite{Mueller2013}, DOTA-puromycin~\cite{Eiger2013}, DOTA dimeric-cyclic Arg-Gly Asp (RDG) peptide~\cite{Hernandez2014}, DTPA Cetuximab Fab~\cite{Chakravarty2014} and DOTA-hypoxia-associated PET tracers~\cite{Szucs2022}. However, most publications (22) have been devoted to the $^{44}$Sc-PSMA-617 radiobioconjugate used for prostate cancer imaging. Although many preclinical studies have been conducted, $^{44}$Sc radioconjugates have only been used twice in human studies. In these studies, two patients received $^{44}$Sc-octreotide~\cite{Singh2017}, and a total of five patients received $^{44}$Sc-PSMA-617~\cite{Eppard2017, Khawar2018}.

Another potential advantage of $^{44}$Sc versus $^{68}$Ga in nuclear medicine is its potential use in theranostic pair with the therapeutic $\beta^-$ emitter $^{47}$Sc~\cite{Choinski2021}. So, $^{44}$Sc and $^{47}$Sc constitute a matched pair of radioisotopes for imaging as well as cancer therapy. It's important to note that the similar structure of Sc-DOTA and Lu-DOTA complexes and the identical lipophilicity of radioconjugates labeled with $^{44}$Sc and $^{177}$Lu enable the use of $^{44}$Sc in theranostic pair not only with $^{47}$Sc but also with the widely-used $^{177}$Lu~\cite{Majkowska2011}.

\subsection*{$^{55}$Co radionuclide}

In the case of larger biomolecules with pharmacokinetics in the range of 1--2 days, such as affibody and nanobody, there are no radionuclides as favourable for $\beta^+\gamma$ coincidence techniques as $^{44}$Sc is for peptides. Among the proposed intermediate-lived positron-emitting radionuclides, $^{55}$Co ($t_{1/2} = 17.53$ h) best meets these conditions, emitting $\beta^+$ in 77\% and $\gamma$ quanta 931.1 keV in 75\%. Practically, two reactions are best suited for producing $^{55}$Co: $^{58}$Ni(p,$\alpha$)$^{55}$Co and $^{54}$Fe(d,n)$^{55}$Co, which are accessible on medium-energy cyclotrons. Using a $^{58}$Ni isotopically enriched target has a significant advantage because its price is much lower than $^{54}$Fe. However, in the competing reactions (p,2p) and (p,p+n), the significant impurity $^{57}$Co is co-produced directly or by $^{57}$Ni. The second $^{54}$Fe(d,n) reaction offers a very high (99.995\%) radionuclide purity product, where purities practically are limited by the achievable enrichment of the $^{54}$Fe target. Small impurities of $^{52\text{m}}$Mn formed in $^{54}$Fe(d,$\alpha$)$^{52\text{m}}$Mn are easily removed during the separation process, where more than 99.99\% of the co-produced $^{52\text{m}}$Mn is removed from the $^{55}$Co. In this method, the high cost of enriched $^{54}$Fe requires recycling. The radiochemical separation of $^{55}$Co from iron targets was performed in HCl solution utilizing various distribution coefficients of cobalt and iron on the anion exchange resins and the extraction DGA resin (N,N,N',N'-tetrakis-2-ethylhexyldiglycolamide) as a step for removal of Mn radionuclides~\cite{Valdovinos2017}. The best results of $^{55}$Co separation were obtained by using DGA resin where recovery was more than 95\% and $^{55}$Co was eluted in $\leq$ 600~$\mu$L fraction. The recovery of the expensive $^{54}$Fe target exceeds 94\%.

The DOTA and NOTA ligands are frequently used to bind $^{55}$Co with biomolecules. NOTA can be radiolabeled with Co at room temperature, while DOTA usually requires additional heating. It's worth noting that DOTA forms weaker complexes with Co$^{2+}$ compared to, for example, its complexes with Sc$^{3+}$ ($K_{\text{Co}^{2+}\text{(DOTA)}} = 19.1$)~\cite{Martell2004}. Due to the high kinetic stability of the complexes, no release of $^{55}$Co is observed from the tested radiobioconjugates.

Since its introduction in 1996~\cite{Jansen1996}, there has been extensive research on the medical applications of $^{55}$Co in positron emission tomography. To date, 156 papers have been published on this subject. Initially, free cobalt in the form of $^{55}$Co$^{2+}$ was used in neurology, allowing to monitor Ca$^{2+}$ behaviour and helping to localise brain ischemia due to their accumulation in damaged tissue~\cite{Stevens1997, Jansen1998, DeReuck1999}. However, further studies have shown that other modalities, especially MRI, are better for the brain imaging~\cite{Stevens1999}. Most of the studies conducted using $^{55}$Co focus on radiolabeled peptides and small biologically active molecules, similar to those of $^{44}$Sc. Cellular and biodistribution studies in mouse models have shown that cobalt-labeled somatostatin receptor-targeted agents have the highest affinity ever observed for somatostatin receptor 2 (SSTR2)~\cite{Heppeler2008}. Also, DOTATATE labeled with $^{55}$Co shows improved PET imaging in comparison to DOTATATE labeled with $^{68}$Ga and $^{64}$Cu~\cite{Andersen2020}. Several publications have been also devoted to the $^{55}$Co-PSMA-617 radio-conjugate used for prostate cancer imaging. The binding to LNCaP and PC3-PIP cell $^{68}$Ga, $^{111}$In, and $^{55}$Co PSMA-617 conjugates were compared. The observed $K_D$ values were similar~\cite{Dam2017}. $^{55}$Co-labeled peptide bombesin have been shown to image PC3 prostate cancers successfully~\cite{Mitran2017}. Also, $^{55}$Co has been used to label the gastrin-releasing peptide receptor (GRPR). GRPR is overexpressed by a large number of cancers including prostate, pancreatic, colon, and non-small cell lung cancers. Neurotensin receptor imaging has emerged as a promising target for imaging pancreatic, colorectal, and breast cancers. Houson~\textit{et~al.}~\cite{Houson2022} examined imaging of overexpression of this receptor using a chelator coupled with neurotensin peptide analogue (NOTA-NT-20.3), a 13-amino-acid peptide with high affinity to neurotensin receptors. Among the tested peptides labelled with positron-emitting radiometals, $^{55}$Co-NOTA-NT demonstrated superior targeting of neurotensin receptors compared to $^{68}$Ga and $^{64}$Cu-NOTA-NT.

The longer half-life of $^{55}$Co, compared to $^{44}$Sc and $^{68}$Ga, makes it suitable for labeling also larger biomolecules with slower pharmacokinetics. For instance, it can be used to label fragments of monoclonal antibodies, such as affibodies and nanobodies, which are part of heavy chain antibodies with a molecular weight of only 12--15~kDa. Affibodies and nanobodies selectively bind to a specific antigen, such as the human epidermal growth factor receptor 2 (HER2), which is overexpressed in approximately 20\% to 30\% of breast cancer tumors as well as in certain ovarian, pancreatic, and gastric cancers. It is linked to more aggressive disease, higher recurrence rates, and increased mortality. The monoclonal antibody trastuzumab is a HER2 receptor blocker, therefore, after labeling with longer-lived radioisotopes, such as $^{89}$Zr ($t_{1/2}=3.3$~d) or $^{124}$I ($t_{1/2}=4.2$~d), it is used in PET diagnosis of these tumors~\cite{Dijkers2009, Orlova2009}. Also, the HER3 protein is expressed in several types of cancer. This fact, together with the role of HER3 in rapid cell proliferation, resistance to drugs targeting other HER receptors, and resistance to some chemotherapeutic drugs, indicates the need for diagnostics of these receptors. However, when using a whole antibody such as trastuzumab, blood clearance is delayed and imaging is usually performed 4--5 days after radiotracer injection~\cite{Dijkers2010}. Using an antibody fragment, such as a nanobody or affibody, for imaging enables faster radioconjugate accumulation in the tumor and faster clearance, usually \textit{via} the kidneys. Such radioconjugates enable imaging on the same day or the next day. The most suitable for this purpose are radionuclides with a $t_{1/2}$ of about one day, and this requirement is ideally met by $^{55}$Co. Additionally, it can be used in the $\beta^+\gamma$ coincidence PET technique. So far, several studies have been conducted on the diagnostics of HER2 and HER3 receptors using affibodies labeled with cobalt radionuclides $^{55}$Co or $^{57}$Co as a $^{55}$Co surrogate~\cite{Rinne2020, Rosestedt2017, Barrett2021}. It was found that the anti-HER2 [$^{57}$Co]Co-DOTA-Z2395-C and anti-HER3 [$^{57}$Co]Co-ZHER3 affibodies showed promising tumor uptake, rapid blood clearance, and good imaging properties in the models tested. It can be expected that after $^{55}$Co labeling, the studied nanobodies will find application in PET imaging~\cite{Barrett2021}.

Research has explored the potential use of $^{55}$Co for imaging the epidermal growth factor receptor (EGFR) overexpression. EGFR is a protein in cell membranes that plays a role in cell growth and replication, and its overexpression can indicate resistance to chemotherapy. The EGFR-targeting affibody ZEGFR:2377 was labeled with $^{111}$In, $^{68}$Ga, and $^{57}$Co as a substitute for $^{55}$Co~\cite{Garousi2017}. In a mouse model study, [$^{57}$Co]Co-ZEGFR:2377 had higher uptake in tumors and lower uptake in the liver and spleen compared to $^{111}$In and $^{68}$Ga labeled ZEGFR:2377. The reason for this disparity was attributed to the presence of a negative charge in the Co$^{2+}$ complexes with DOTA, as opposed to the neutral charge found in the In$^{3+}$ and Ga$^{3+}$ DOTA complexes.

\subsection*{$^{124}$I radionuclide}

Monoclonal antibodies have revolutionized cancer therapy by providing targeted treatments that specifically attack cancer cells while minimizing damage to healthy tissues. Their therapeutic use is extensive, e.g., in HER2+ breast cancers, where trastuzumab and its conjugates with chemotherapeutics such as Kadcyla or Enhertu are the main therapeutic agents. However, the use of antibodies in PET diagnostics is very limited due to the lack of $\beta^+$ emitters with an appropriately long half-life and the need for stable labeling of antibodies at room temperature. Radionuclide $^{124}$I is an alternative long-lived PET radioisotope attracting increasing interest for long-term clinical and PET studies~\cite{Cascini2014}. Since $^{124}$I emits both $\beta^+$ and prompt gamma it can still be utilized in the $\beta^+\gamma$ coincidence technique (Table 1). The best method of producing $^{124}$I is the $^{124}$Te(p,n) reaction~\cite{Aslam2010, Bzowski2022} which can be used in small medical cyclotrons. Recent measurements of the reaction efficiency indicate that irradiation of a $^{124}$Te target in the energy range of 14$\rightarrow$7 MeV at a current of 20~$\mu$A allows the production of a few hours of 50~MBq batches of $^{124}$I used for imaging~\cite{Plyku2017}. The classical dry distillation method separates $^{124}$I from the tellurium target, allowing easy recycling of the enriched tellurium without further handling.

The $^{124}$I itself is extremely valuable for imaging thyroid cancer because it can be taken up by the thyroid gland in a similar way to $^{131}$I, which is commonly used in thyroid therapy. For imaging other cancers, it is necessary to attach $^{124}$I to the target ligand, which in the case of iodine is simple due to its chemical properties and ability to form covalent bonds with proteins -- antibodies. An example of a clinical trial using $^{124}$I is $^{124}$I-labeled A33 in PET imaging to assess tumour location and to help guide therapy~\cite{Zanzonico2015}. Another monoclonal antibody, Cetuximab, targets the epidermal growth factor receptor (EGFR), which is overexpressed in several cancers, including colorectal and head-and-neck cancers. PET imaging with $^{124}$I-labeled Cetuximab is helpful in selecting patients for therapy and monitoring response to treatment~\cite{Spiegelberg2016}. The full-length trastuzumab antibody targets the HER2 receptor. PET imaging with $^{124}$I-trastuzumab is used to assess HER2 expression in tumors and metastatic sites of metastatic breast cancer (MBC), and to assess the distribution of the therapeutic antibody. Results demonstrate that $^{124}$I-trastuzumab effectively detects HER2-positive lesions in primary and metastatic cancer patients and can quantitatively differentiate HER2-positive and HER2-negative lesions~\cite{Guo2020}. In HER2-negative patients with MBC the $^{124}$I-labelled epichaperome inhibitor (PU-H71) was used as a non-invasive predictive biomarker, indicating a new diagnostic and therapeutic option in these patients in combination with nab-paclitaxel~\cite{Jhaveri2020}. Among the $^{124}$I-labeled radioligands, [$^{124}$I]I-MIBG (M-iodobenzylguanidine) has been used for clinical imaging of neuroblastoma and pheochromocytoma~\cite{Hartung2012, Ficola2013}.

In summary, we can conclude radionuclides such as $^{44}$Sc, $^{55}$Co, and $^{124}$I are suitable for labeling commonly used biological vectors and newly developed ligands in PET. Their properties allow for their use in $\beta^+\gamma$ coincidence for positronium imaging as a new biomarker with the potential to distinguish healthy tissue from cancer tissue according to their physical and biochemical properties as was defined in previous studies using $^{68}$Ga (myxoma vs. adipose tissue, brain vs. glioblastoma)~\cite{Moskal2021d, Moskal2024a}. Routine PET imaging with new radiopharmaceuticals listed above, even if there are new radioisotopes emitting additional prompt gamma, provide answers only to the questions: (1) what is the accumulation of the radiopharmaceutical in the imaged or studied tissue or (2) what are the kinetics of the radiopharmaceutical in the imaged organ~\cite{Gu2023}. Such imaging does not allow determining the stage of advancement of the tumour in situ, for this purpose it is necessary to perform a biopsy. Gaining information about the molecular structure of the lesion will be possible by the application of the oPs lifetime, which as an additional imaging parameter with respect to the metabolic characteristics (e.g. FDG) or assessment of the density of surface receptors (e.g. HER2, PSMA). We anticipate that PET imaging enhanced with positronium imaging holds promise for obtaining a virtual biopsy, providing metabolic, molecular, and structural reconstruction of tissue.

\section{First direct positronium lifetime \\images of humans}

The first \textit{in vivo} positronium lifetime images have been demonstrated for the patient with the secondary recurrent glioblastoma~\cite{Moskal2024a}. The studies were conducted in the Medical University of Warsaw in Poland using the portable multi-photon J-PET scanner constructed at the Jagiellonian University in Cracow, Poland. The photograph of the patient with the head inside the J-PET scanner is shown in Fig.~\ref{fig:jpet_decay}. The patient was treated with alpha particles emitted in the decay of $^{225}$Ac radionuclide delivered intratumorally by administration of the 20~MBq of [$^{225}$Ac]Ac-DOTAGA-SP together with 8~MBq of [$^{68}$Ga]Ga-DOTA-SP. $^{68}$Ga radionuclide was administered to monitor the accumulation sites where the $^{225}$Ac is delivered by Substance-P. Substance P possesses high affinity to the neurokinin-1 receptors overexpressed on the glioblastoma cells surface and within the tumor neovasculature~\cite{Krolicki2020}. Moreover, 131 minutes before the treatment, the patient was injected intravenously a 178~MBq activity of [$^{68}$Ga]Ga-PSMA-11 for treatment planning purposes. The PSMA-11 molecule was used since it also shows high specificity for glioma by binding to receptors on microvascular endothelial cells in the glioma tumor.

The imaging with J-PET scanner was performed twice: once about 90 minutes after injection of [$^{68}$Ga]Ga-PSMA-11, and then second time about 16 minutes after intratumoral administration of [$^{225}$Ac]Ac-DOTAGA-SP and [$^{68}$Ga]Ga-DOTA-SP. In both cases, J-PET was used after completion of regular PET diagnostics with Biograph 64 TruePoint PET/CT.

\begin{figure}[htbp]
	\centering
	\includegraphics[width=\linewidth]{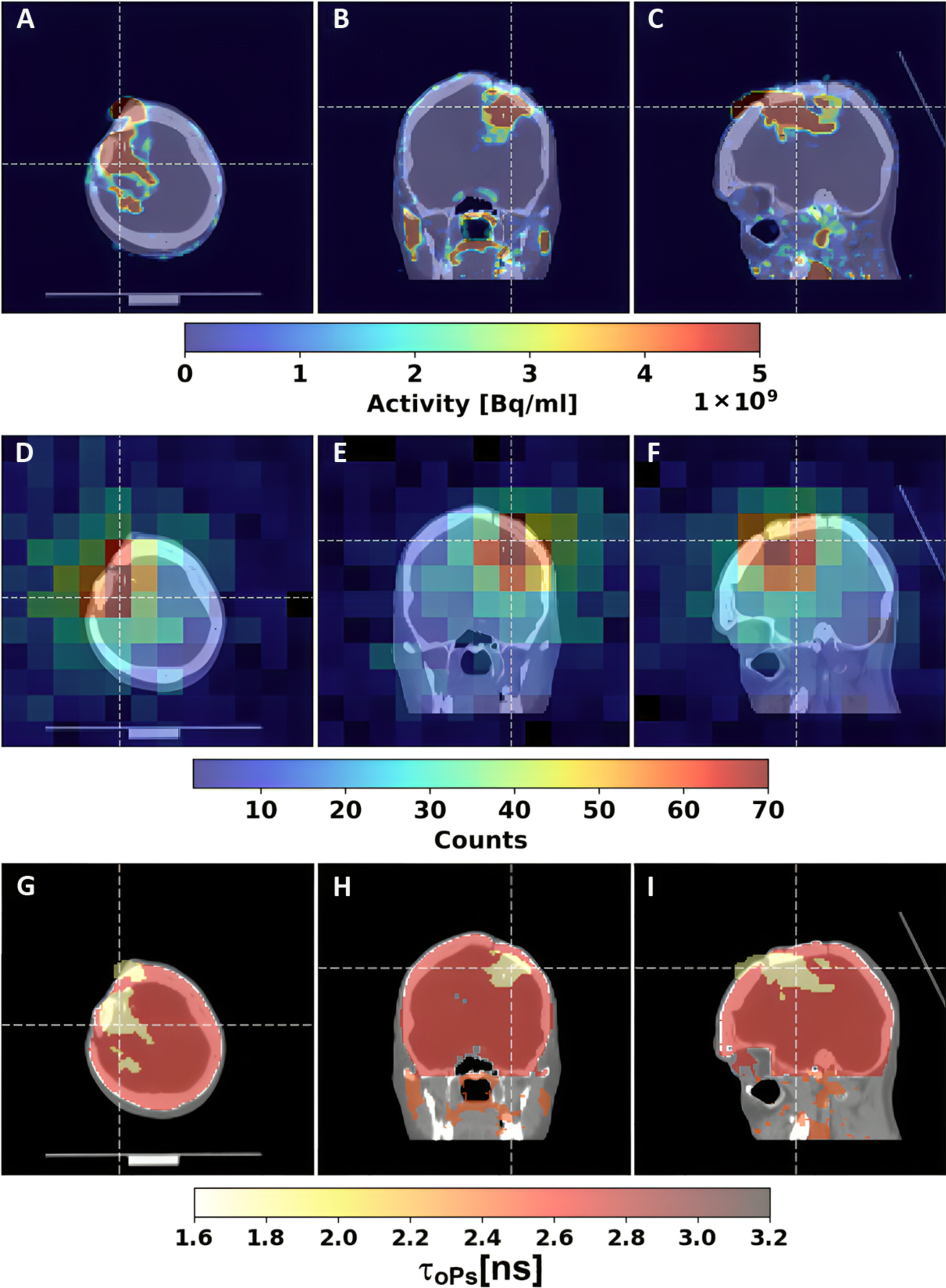}
	\caption{
		Standard 2$\gamma$ PET images, $\beta^+$+$\gamma$ coincidence PET images, and first positronium lifetime images of the head of a patient with recurrent secondary glioblastoma in the right frontoparietal lobe~\cite{Moskal2024a}.
		Images were acquired after intracavitary injection of [${}^{68}$Ga]Ga-DOTA-SP together with [${}^{225}$Ac]Ac-DOTA-SP radiopharmaceuticals.
		Transverse (left column), coronal (middle column), and sagittal (right column) cross sections are presented.
		White dashed lines indicate cross-section planes corresponding to the other two orientations.
		Upper row (A, B, C) presents standard PET/CT fusion images showing accumulation of [${}^{68}$Ga]Ga-DOTA-SP obtained with the Biograph 64 TruePoint scanner.
		Middle row (D, E, F) presents the first in-human $\beta^+$+$\gamma$ coincidence images acquired using the multi-photon J-PET system.
		Lower row (G, H, I) shows the first \textit{in vivo} positronium lifetime images in humans, presenting the mean lifetime of ortho-positronium ($\tau_{\mathrm{oPs}}$).
		The color scale of $\tau_{\mathrm{oPs}}$ values is shown below the figure.
		Standard PET images were used to delineate glioma tumor, brain, and salivary gland regions for which mean positronium lifetimes were evaluated.
		Figure and caption adapted from~\cite{Moskal2024a}.
		\textcopyright~2024 American Association for the Advancement of Science.
		Reprinted with permission from Moskal~\textit{et~al.}~\cite{Moskal2024a} under a Creative Commons Attribution License 4.0 (CC BY).
	}
	\label{fig:pet_images}
\end{figure}

Upper panel of Fig.~\ref{fig:pet_images} shows the standard 2$\gamma$ PET images obtained with Biograph 64 TruePoint PET/CT. The images demonstrate the highest accumulation of the pharmaceutical [$^{68}$Ga]Ga-DOTA-SP in the tumor cavity located in the right frontoparietal lobe, and also show high accumulation of PSMA-11 molecule in the salivary glands.

Middle panel of Fig.~\ref{fig:pet_images} shows distributions of reconstructed annihilation places for 2$\gamma$ annihilations provided that also a prompt gamma ($\gamma_{\text{prompt}}$) from the events where $^{68}$Ga decay was registered. The obtained statistics is very low because for $^{68}$Ga radionuclide the decay \textit{via} process $^{68}$Ga~$\rightarrow$~$^{68}$Zn* + $e^+$ + $\nu$~$\rightarrow$~$^{68}$Zn + $\gamma_{\text{prompt}}$ + $e^+$ + $\nu$ constitutes only $Y_{\beta^++\gamma}/Y_{\beta^+} = 1.34\%$ (Table~I), and because the requirement of registration of prompt gamma in addition to annihilation photons lowered the detection efficiency of the J-PET prototype to only 0.06~cps/kBq~\cite{Moskal2024a}. The low statistics in this first trial with $^{68}$Ga was expected, therefore, the experiment was conceived for the brain with the local administration of pharmaceutical to maximize the signal to background ratio and to minimize the attenuation and scatterings of photons in the body. J-PET scanner with the triggerless data acquisition enable to register two, three and more photons simultaneously. The annihilation and prompt photons are disentangled based on the energy deposited in scintillators \textit{via} Compton effect~\cite{Moskal2014, Moskal2024a}. The maximum energy deposition expected for various $\beta^+\gamma$ radionuclides is presented as Comp.~E. in Table~I.

The registered 2$\gamma+\gamma_{\text{prompt}}$ events were used to determine the lifetime of ortho-positronium ($\tau_{\text{oPs}}$), and also the mean positron annihilation lifetime ($\tau_{e^+}$). The first \textit{in vivo} images of the $\tau_{\text{oPs}}$ values in the brain are shown in the lower panel in Fig.~\ref{fig:pet_images}. The values for $\tau_{e^+}$ and $\tau_{\text{oPs}}$ are also presented in Fig.~\ref{fig:oPs_in_vivo}. The determined value of $\tau_{\text{oPs}} = 1.77 \pm 0.58$~ns in glioma tumor is smaller than $\tau_{\text{oPs}} = 2.44 \pm 0.46$~ns in salivary glands, which in turn is smaller than $\tau_{\text{oPs}} = 2.72 \pm 0.72$~ns in healthy brain tissues~\cite{Moskal2024a}. The same tendency but with higher statistical significance is observed for mean positron lifetime with $\tau_{e^+} = 1.28 \pm 0.06$~ns for the glioma tumor, 1.53~$\pm$~0.04~ns for salivary glands, and 1.89~$\pm$~0.06~ns for healthy brain tissues~\cite{Moskal2024a}.

This result demonstrated for the first time the feasibility of performing positronium lifetime images in clinics and revealed that there are substantial differences in positron and ortho-positronium lifetime in glioblastoma cells, salivary glands and those in healthy brain tissues, indicating that positronium imaging could be used to diagnose disease \textit{in vivo}~\cite{Moskal2024a}. In order to translate the positronium lifetime imaging into routine clinical practice the development of more efficient multi-photon PET scanners is required~\cite{Moskal2021e, Baran2025} or adaptation to multi-photon mode of the current long axial field-of-view PET scanners~\cite{Steinberger2024, Mercolli2024}. Moreover, the development of the iterative methods for positronium lifetime imaging is also needed to improve the spatial and temporal resolutions of the direct images limited by the time-of-flight resolution of the current PET scanners.

\section{Iterative reconstruction methods \\for positronium lifetime imaging}

\subsection*{Reconstruction-less methods}

Traditionally, positronium lifetime images are formed in two separate steps: localizing the annihilation point event-by-event and estimating the lifetime for each voxel by fitting its delay-time histogram. In 2 annihilations, the annihilation point can be estimated using the time-of-flight (TOF) information and each event is assigned to its most likely position~\cite{Moskal2019b}. This method has been used to create the first positronium lifetime image~\cite{Moskal2021d}, but the spatial resolution is largely limited by the TOF resolution, except in two-dimensional imaging where the activity locates only on a plane that is perpendicular to the lines of response~\cite{Takyu2022}. In current state-of-the-art PET scanners with a TOF resolution of 200--500~ps, the estimated spatial resolution is about 30--75~mm~\cite{Moskal2020a}. A modified method to improve the lifetime image quality assigns a weighted time delay to multiple voxels based on the probability derived from an estimated activity distribution~\cite{Shopa2022,Shopa2023}. This method reduces noise but does not improve the spatial resolution. One way to improve the spatial resolution in the reconstruction-less approach is using whole-gamma imaging that combines PET and Compton imaging~\cite{Yamaya2024}. The Compton cone of the prompt gamma can be used to reduce the localization uncertainty. As an alternative, positronium lifetime images can also be formed using 3 annihilations, where the annihilation point of each event can be estimated using the detection time and energy information of the three detected photons~\cite{Abuelhia2007,Kacperski2004,Kacperski2005} or time and position of the detected photons~\cite{Moskal2019b, Moskal2021c}. However, the occurrence of 3 annihilations in biological tissues is less than 1\% of 2 annihilations, making it more difficult to achieve sufficient counts for lifetime image estimation.

Once all the events are localized, oPs lifetime inside each voxel can be estimated by fitting the time-delay histogram to a lifetime model. A commonly used model is a summation of multiple exponential functions convolved with a Gaussian function~\cite{Kansy1996,Olsen2022}:

\begin{equation}
	f(t) = \sum_{i} I_i \cdot e^{-\lambda_i (t - \delta)} \cdot H(t - \delta) \ast G(t;\delta,\sigma) + B,
	\label{eq:emg}
\end{equation}

\noindent
where $g(\tau; \mu, \sigma)$ is a Gaussian function with mean $\mu$ accounting the detector timing offset and standard deviation $\sigma$ accounting for the detector timing resolution and the residual errors in correcting the travel time of the prompt gamma and the annihilation photon pair before detection. The subscript $l \in \{o, p, d\}$ denotes an annihilation pathway corresponding to oPs, pPs and direct annihilation, respectively. $A_l$ denotes the intensity of the $l^\text{th}$ pathway with $A_o + A_p + A_d = 1$ and $\lambda_l$ is the annihilation rate (i.e., the inverse of the lifetime). $u(t)$ is the unit step function and $B$ models the flat background. The above convolution between an exponential function and a Gaussian function is also known as exponentially modified Gaussian (EMG). Depending on the signal composition, fewer or more components may be used. The fitting procedure can be carried out using existing software developed for PALS~\cite{Dulski2020,Olsen2022} or a customized optimization algorithm~\cite{Shopa2023,Berens2024}. Alternatively, the lifetime can also be estimated using the inverse Laplace transform~\cite{Shibuya2022}, instead of least squares fitting.

The above-mentioned methods are referred to as \textit{reconstruction-less methods} because event positioning is carried out one event at a time. These methods are simple to implement but have poor spatial resolution. Below we will review recently developed methods for high-resolution lifetime image reconstruction that include direct and indirect reconstruction methods.

\subsection*{Direct maximum likelihood reconstruction}

The first effort to overcome the resolution limited set by the TOF resolution is the development of a direct maximum likelihood (ML) reconstruction method by Qi and Huang~\cite{Qi2022}. The direct ML method performs the two tasks, event localization and lifetime estimation, simultaneously by maximizing the likelihood function of the lifetime events. It can improve the spatial resolution of lifetime image to the same level of regular PET images by incorporating the PET system response function in positronium lifetime image reconstruction~\cite{Qi2022}. Note that since the goal here is to reconstruct the lifetime image, not the activity image, the system response model does not need to consider the prompt gamma detection efficiency. This greatly simplifies the computation and allows the use of standard PET system matrix in lifetime image reconstruction. Based on a mono-exponential lifetime model, Qi and Huang derived an iterative algorithm using the optimization transfer principle to find the ML estimate of the positronium lifetime image. The algorithm has a closed-form update equation that resembles the MLEM algorithm for emission tomography. It also shares similar properties as those of the MLEM algorithm, including monotonically increasing the log likelihood function and guarantee of non-negativity. To reduce image noise, they further developed a penalized ML (PML) reconstruction algorithm using pairwise L1 and L2 penalty functions. Again, an iterative algorithm with a closed-form update equation was obtained to maximize the penalized likelihood function. Simulation studies showed that the ML and PML methods can recover high-resolution lifetime images using existing TOF PET technologies.

Since the mono-exponential model is too simplistic for positronium lifetime events, Huang~\textit{et~al.}~\cite{Huang2022} proposed a PML method using the EMG model, i.e., only one lifetime event in (1). Instead of deriving a closed-form update equation, they used a standard optimization solver, Limited-memory Broyden-Fletcher-Goldfarb-Shanno Bound (L-BFGS-B) method, available in \textit{scipy}. One advantage of using a standard solver is that the method can be easily extended to include two lifetime components without additional derivation~\cite{Chen2023,Chen2024}. Simulation results demonstrated that the EMG model outperforms the mono-exponential model for realistic data.

\subsection*{Indirect reconstruction methods}

\begin{figure}[htbp]
	\centering
	\includegraphics[width=\linewidth]{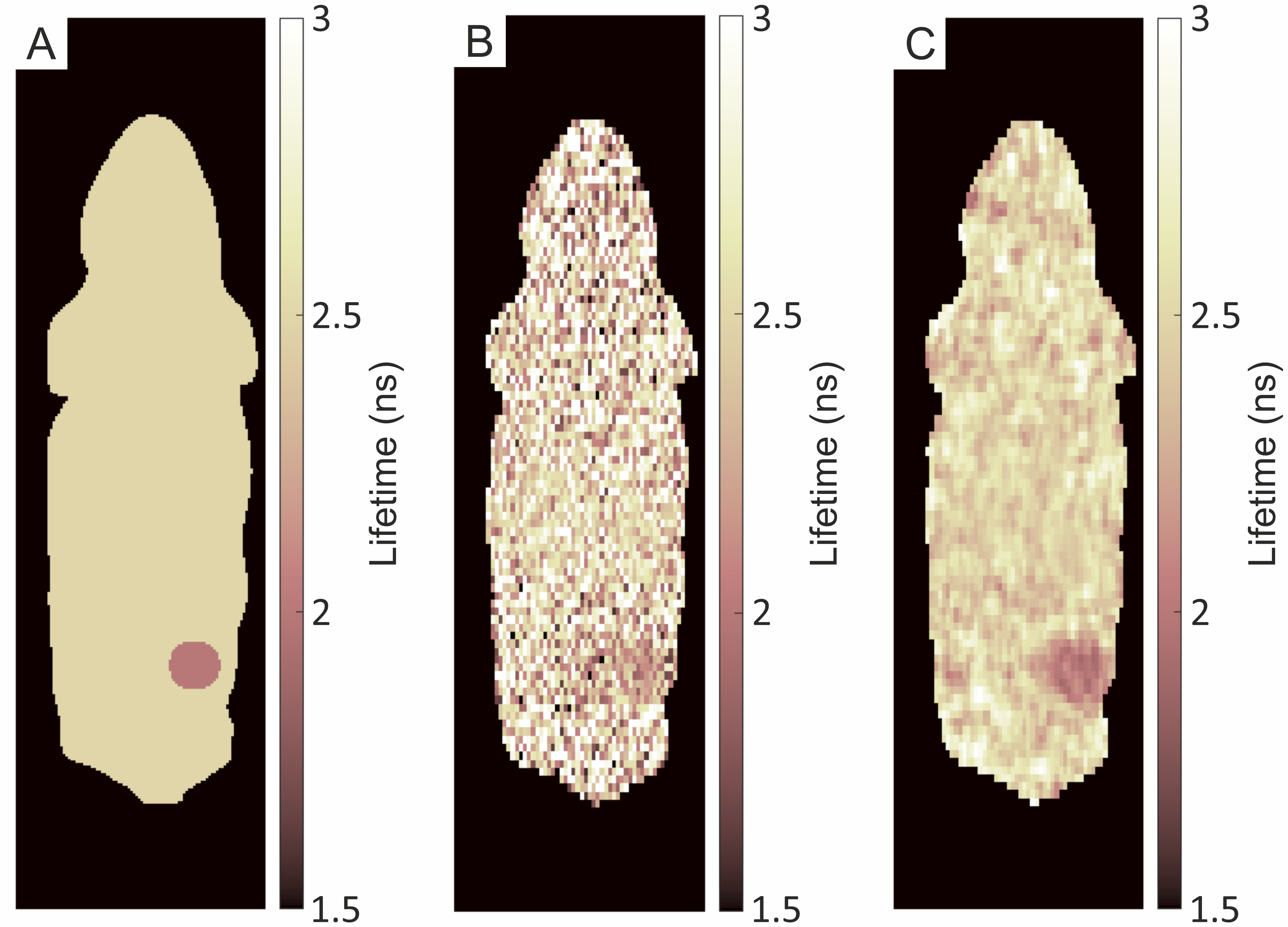}
	\caption{
		Comparison of reconstructed ortho-positronium ($\mathrm{oPs}$) lifetime images of a simulated 3D rodent phantom:
		(A) the ground truth lifetime image;
		(B) reconstruction using the direct TOF method;
		(C) reconstruction using the SPLIT method.
		More details can be found in~\cite{Huang2024b}.
	}
	\label{fig:phantom_simulation}
\end{figure}

One of the drawbacks of the above direct ML methods is their high computation cost. This was the reason that only 2D simulation studies were conducted in~\cite{Qi2022,Huang2022,Chen2023,Chen2024,Huang2025a}. To overcome the computational challenge and be applicable to more-than-two-components models, Huang~\textit{et~al.}~\cite{Huang2024b} proposed a new positronium lifetime reconstruction method based on time thresholding, which was referred to as SPLIT (Statistical Positronium Lifetime Image reconstruction by Time Thresholding). SPLIT is an indirect lifetime image reconstruction method because it decouples the activity image reconstruction and lifetime estimation by first reconstructing a sequence of lifetime-encoded activity images and then using the reconstructed images to estimate the lifetime pixel-by-pixel. The lifetime-encoded images are obtained by assigning a set of lifetime-dependent weighting factors to each event. A simple weighting scheme given in~\cite{Huang2024b} uses binary weight based on a threshold. An event is assigned either a weight of 1 (kept) if the time delay between the annihilation photons and the prompt gamma is less than a preset threshold $T_c$, or a weight of 0 (removed) otherwise. The weighted data were reconstructed by the ordered subset expectation maximization (OSEM) algorithm to produce a lifetime-encoded image. Using a series of pre-determined threshold $T_c$’s, a time-series of images can be obtained that can then be used to estimate a positronium lifetime image voxel-by-voxel. There are two advantages of the SPLIT method: the first one is the use of the existing OSEM reconstruction algorithm, so it is computationally fast and can be easily adapted to different PET scanners; the second one is that it can adopt any lifetime model, like the one in (1), because the reconstruction and lifetime estimation are decoupled. Because of its computational efficiency, the SPLIT has allowed reconstruction of positronium lifetime image from fully 3D dataset. As shown in Fig.~\ref{fig:phantom_simulation}, the SPLIT method can reconstruct oPs lifetime images with higher resolution and less noise, than the direct TOF method.

Another indirect lifetime reconstruction method is the \nobreak{SIMPLE} (Statistical IMage reconstruction of Positronium Lifetime by time wEighting) method~\cite{Huang2022b,Huang2025b}, which weights each event by its time delay directly. The SIMPLE method provides an efficient approach to the reconstruction of the average lifetime image using two OSEM-type reconstructions: one is reconstruction of an activity image using 511-keV photons in the triple coincidence events and the other is reconstruction of a lifetime weighted image. The first reconstruction is a standard OSEM reconstruction. The second reconstruction is a modified list-mode OSEM algorithm by weighting each event by its time delay measurement. The averaged lifetime image is then obtained by taking the ratio between the time-weighted image and the activity image. The SIMPLE method is extremely efficient in computation because it only requires one additional reconstruction of a time-weighted dataset using the OSEM algorithm. It does not require any curve fitting and is quite robust for average lifetime estimation. While only the oPs lifetime is of interest in positronium lifetime imaging, the average lifetime of all annihilation pathways can also show tissue contrast and in some case, may offer higher contrast-to-noise ratio as shown in~\cite{Moskal2024a}. Most recently, a regularized SIMPLE method based on the kernel method has also been developed to reduce noise in the estimate lifetime image~\cite{Huang2024c}.

To estimate the oPs lifetime image, Huang and Qi extended the SIMPLE method to include additional weighting factor of higher-order moments of the time delay, i.e., $\tau^n$, where $n$ is the order of the moment~\cite{Huang2024d}.

The extension enables an estimation of the oPs lifetime in each voxel using the method of moments, while preserving the computational efficiency. In general, for a lifetime model with $N$ unknown parameters, at least $N$ images weighted by $\tau, \dots, \tau^N$, respectively, are needed. Note that the timing offset $\mu$ and uncertainty $\sigma$ can be estimated from the global time delay histogram before image reconstruction, so the real unknown parameters in the lifetime image reconstruction are intensities $A_i$’s and rates $\lambda_i$’s in (1). By setting $\lambda_d$ and $\lambda_p$ to their known values and using the fact $\frac{A_o}{A_p} = 3$, only the first- and second-moment weighted images are needed to calculate the oPs lifetime image.

\subsection*{Data corrections}
\underline{Travel time correction}: Travel times of the three photons in a lifetime event need to be corrected before calculating the lifetime measurement $\tau$. For this correction, an annihilation point is inferred as the most likely position along an LOR by the TOF information of the two 511~keV photons. The emission time of each photon is thus estimated by subtracting the travel time from the estimated annihilation point to its detection point from its original detection time. The time measurement of each lifetime event is therefore $\tau = \hat{t}_{511} - \hat{t}_{PG}$, where $\hat{t}_{511}$ and $\hat{t}_{PG}$ are the estimated emission times of the 511-keV photons and the prompt~$\gamma$, respectively. Error in travel time correction is modeled in the Gaussian function $g(\tau; \mu, \sigma)$ in Eq.~(1).

\underline{Random events estimation and correction}: The accuracy of lifetime estimates can be compromised by random events if left uncorrected. Here we briefly review the classification and estimation of randoms presented in~\cite{Huang2024b}. Based on the criterion whether every two single events in a triple coincidence are from the same decay, the random events can be categorized into three types. Type~I randoms are formed by a pair of 511-keV photons from the same annihilation with a random prompt gamma; Type~II randoms are formed by a 511-keV photon and a prompt gamma from the same decay with a random 511-keV photon; Type~III randoms are formed by photons from three different decays. Similar to the random estimation in PET, randoms in positronium lifetime imaging can be estimated using delayed windows. Four types of delayed windows were designed in~\cite{Huang2024b} to estimate the three types of randoms. Alternatively, the randoms rate can be estimated from count rates of prompt gammas and 511-keV photons. Once estimated, types~II and III randoms can be corrected as a background events in the reconstruction step, whereas type~I randoms can be corrected in the image domain during the lifetime fitting procedure.

\section{Developing methods for positronium lifetime validation with clinical scanners}

Positronium (Ps) lifetime spectroscopy has the potential to add an additional imaging modality to clinical positron emission tomography (PET) scans. This new modality could yield a greater degree of diagnostic information because the lifetime of Ps is determined by the surrounding void structure and chemical composition. Not only could a lesion be identified based on uptake, but characteristics of the lesion such as hypoxia~\cite{Shibuya2020} may be identifiable. In theory, general uptake in organs could also be used to identify the health of those organs~\cite{Jasinska2017c} or identify tissue differences that would not appear in a conventional PET scan. P. Moskal~\textit{et~al.}, demonstrated a significant lifetime difference between tissue from a cardiac myxoma and cardiac adipose tissue~\cite{Moskal2021d}, which suggests that applying Ps lifetime spectroscopy to a [$^{82}$Rb]Cl perfusion scan may be able to identify such a growth in the heart. Patient scans in research scanners such as the J-PET are already underway to begin to understand Ps lifetimes within the human body~\cite{Moskal2024a}. With such research being performed across various scanners with various methodologies it is necessary to implement a kind of standardization to ensure that the same lifetimes can be measured for the same materials independent of the scanner and method. Failure to meet such a standardization would require an independent database of Ps lifetime values for a given scanner or methodology and beg the question: ``Which is actually correct??''

This validation assessment can be further broken down into two subcomponents assuming the scanners and detector systems are yielding accurate response information: the material being measured and the fitting methodology to extract the ortho-Ps (oPs) lifetime. To further complicate matters, these subcomponents are not completely separable. To show that, one can compare results published by K. Shibuya~\textit{et~al.} and P. Stepanov~\textit{et~al.}, which both measured the oPs lifetime of water saturated with air. K. Shibuya~\textit{et~al.} measured the lifetime of air-saturated water to be $1.9040 \pm (0.0029)$~ns~\cite{Shibuya2020} whereas P. Stepanov~\textit{et~al.}, measured the value to be $1.802 \pm (0.015)$~ns~\cite{Stepanov2020}. Both additionally measured that increasing the oxygen concentration of the water decreased the measured oPs lifetime, but the absolute values and the magnitude of the absolute values differ substantially between the two publications. The direct explanation for these differences is that they used two different methodologies to extract the lifetimes. K. Shibuya~\textit{et~al.} fit an exponential function to a specific region of the time-difference distribution (TDD)~\cite{Shibuya2020} while P. Stepanov~\textit{et~al.} fit the lifetimes and intensities in three separate exponential functions to the measured TDDs~\cite{Stepanov2020}. This comparison is not to suggest that one is correct over the other, but to point out that significantly different lifetimes can be measured for the same material depending on the fitting methodology. Moreover, as more \textit{in vivo} patient or tissue measurements are being performed such differences must be understood or standardized to be able to apply Ps lifetime analysis to clinical scans.

Water, although a seemingly obvious choice for a standardized material, has an oPs lifetime that is dependent on temperature~\cite{Kotera2005}, the gasses in solution~\cite{Shibuya2020,Stepanov2020} and the ion concentration in solution~\cite{Green1958}. These factors can be mitigated by using demineralized or distilled water, but may still add a variability to the measurements. A more constant, solid state material would add not only additional calibration points, but would mitigate the latter two effects mentioned with water. The National Metrology Institute of Japan (NMIJ) has begun calibrating and setting measurement standards for material characterization~\cite{NMIJ}. Two such materials for Ps lifetime analysis are quartz glass and polycarbonate, which have referenced lifetimes of $1.62 \pm 0.05$~ns and $2.10 \pm 0.05$~ns respectively~\cite{NMIJ}. These materials were acquired and measured by S. Takyu~\textit{et~al.} who measured the values to be $1.56 \pm 0.08$~ns and $2.07 \pm 0.16$ for the two respective reference materials~\cite{Takyu2022}. These two materials give additional data points with values below and above the oPs lifetime of water allowing for a better assessment of a system’s response for measuring oPs lifetimes.

\begin{figure}[htbp]
	\centering
	\includegraphics[width=\linewidth]{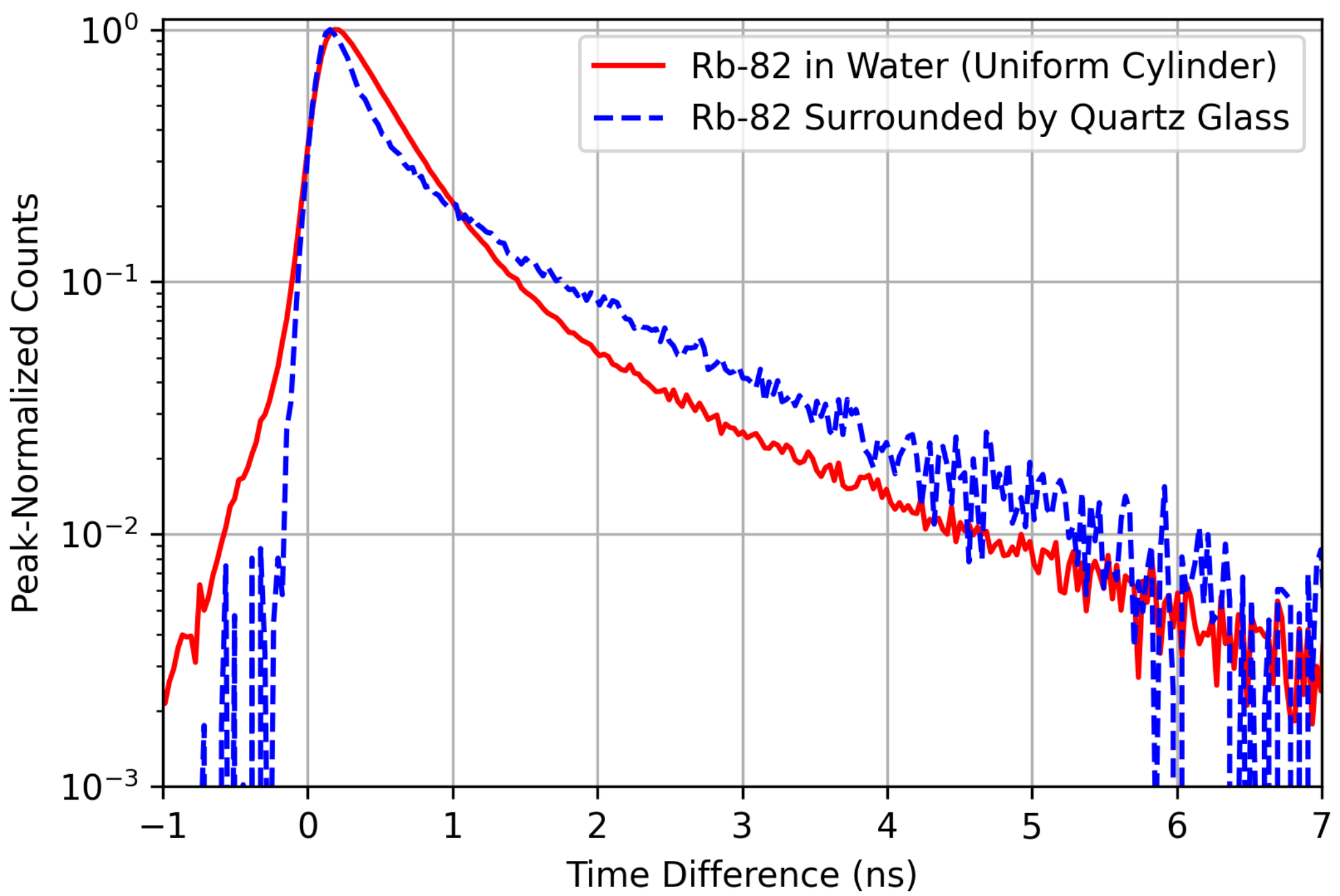}
	\caption{
		Positronium time-difference distributions for quartz glass and water measured using a Siemens Biograph Vision Quadra scanner.
	}
	\label{fig:time_difference_vision}
\end{figure}

To test and validate the response of a clinically available PET scanner, the Ps lifetimes of quartz glass and water samples were measured in a Biograph Vision Quadra (Siemens Healthineers). Recent advances in the acquisition architecture of the Quadra have made it possible to acquire list-mode data composed of single interaction events within a given block detector. This list-mode data can be processed to measure triple-coincident events and generate TDDs, which can be used to extract the oPs lifetime. The quartz-glass samples were chosen as one of the clinically available sources where droplets of the radiotracer solution were placed in between two quartz glass samples. The measured TDDs were fit with the sum of three exponentially-modified Gaussian distributions to represent the components representing the para-Ps (pPs), direct, and ortho-Ps (oPs) lifetimes. A Bayesian method was used to fit the variables of the fit and the following oPs lifetimes were extracted: $1.58 \pm 0.07$~ns using $^{68}$Ga and $1.62 \pm 0.01$~ns using $^{124}$I~\cite{Steinberger2024}. The measured lifetime for the $^{82}$Rb quartz-glass measurement is shown in Fig.~\ref{fig:time_difference_vision}.

These results are all within a single standard deviation of the values published by the NMIJ and those measured by S.~Takyu~\textit{et~al.} The measured oPs results presented in Steinberger~\textit{et~al.} additionally agree with those published by J.~Horn~\textit{et~al.}, who measured the quartz-glass oPs lifetime to be $1.607 \pm 0.06$~ns~\cite{Horn2017}. Horn~\textit{et~al.} also found that the direct annihilation lifetime ($0.156$~ns) and pPs lifetime ($0.524 \pm 0.009$~ns) were significantly longer than the lifetimes that are commonly assumed for fitting the TDDs ($0.125$~ns and $0.388$~ns respectively~\cite{Moskal2021d,Steinberger2024}). The fact that the oPs lifetime values agree well may be due to the fitting methodology that lets the intensities be free parameters in the fit. This result agrees well with Kotera~\textit{et~al.}, who demonstrated that the oPs lifetime only has a slight dependency on the pPs fit intensity~\cite{Kotera2005}. Along with the quartz glass, a water-filled uniform cylinder ($20\,\phi\, \times 30$~cm$^3$) with a $^{82}$Rb solution was also measured in the Quadra scanner. Lifetimes were extracted from various regions of the cylinder to determine if scatter and attenuation impact the shape of the extracted TDDs. Results showed that there was an insignificant shape change between the inner ($5\,\phi\, \times 15$~cm$^3$) and outer ($20\,\phi\, \times 30$~cm$^3$) regions of the cylinder. The overall oPs lifetime of the water was measured to be $1.815 \pm 0.013$~ns~\cite{Steinberger2024}, which also agrees well with the results published by P.~Stepanov~\textit{et~al.} and K.~Kotera~\textit{et~al.} The agreement of both sets of results validate the methodology and capability of the Quadra scanner to perform positronium lifetime imaging and set a firm foundation for future \textit{in vivo} human measurements.

\section{First positronium imaging \\with the clinical scanners}

Following the above-discussed positronium image acquisitions of quartz glass and water samples in the Quadra scanner, the next logical step is to apply this technology to clinical scans. This transition is pivotal for translating experimental findings into practical medical applications. However, it is important to note that only a subset of clinically employed radioisotopes is suitable for positronium imaging. Specifically, radioisotopes such as $^{124}$I, $^{68}$Ga, and $^{82}$Rb qualify due to their properties that facilitate positronium detection. These isotopes are particularly advantageous because of their high uptake in the tumors (either alone or in combination with a ligand), though $^{68}$Ga posseses very low fraction of prompt gamma emission (Table I).

Especially $^{124}$I seems to be a promising candidate for positronium imaging. This is because of its high fraction of prompt gamma emission together with its very high accumulation in thyroid cancer lesions~\cite{Conti2016, Phan2008}. Iodine is needed for the synthesis of thyroid hormones in health thyroid tissue, which is why the natrium iodine symporter (NIS) is strongly expression both on normal thyroid cells and differentiated thyroid cancer~\cite{Chung2002}. The NIS transporter can be utilized both for diagnostic and therapeutic purposes. Depending on the stage and variant of thyroid cancer, patients routinely undergo ablation with $^{131}$I after thyroid cancer resection~\cite{Haugen2016}. Radioiodine effectively eliminates both physiological remnant thyroid tissue and residual tumor cells for improved tumor control. To verify the success of the initial ablation or to detect recurrence of thyroid cancer in case of rising tumor marker levels, iodine imaging can be performed~\cite{Avram2022}. Usually, gamma emitting isotopes like $^{123}$I or $^{131}$I are used for this purpose. However, combined with either planar scintigraphy or single photon emission tomography (SPECT) imaging, they offer only a poor image quality. To overcome those limitations, NIS targeted imaging with $^{124}$I has been proposed, which enables iodine PET imaging which highest resolution~\cite{Phan2008}. Still, thyroid cancer may dedifferentiate over time, which leads to decreasing expression of the NIS transporter and therefore the inability of iodine imaging to detect all lesions~\cite{Grabellus2012}. It is currently not well understood which factors can be used to predict dedifferentiation metastases~\cite{Li2022}. Preliminary data suggest that tumor hypoxia may be involved in the process of cancer dedifferentiation, which leads to more aggressive tumor phenotypes~\cite{Ma2023}. Therefore, positron lifetime imaging could be a tool to investigate the inter lesional heterogeneity of thyroid cancer. The identification of lesions that dedifferentiate seems especially important in the light of approaches that aim for redifferentiation of thyroid cancer, which may restore the ability to accumulate iodine for therapeutic purposes~\cite{Buffet2020}. Prospective pilot trials on $^{124}$I PET for the evaluation of positronium lifetime imaging therefore seem warranted.

\begin{figure}[htbp]
	\centering
	\includegraphics[width=\linewidth]{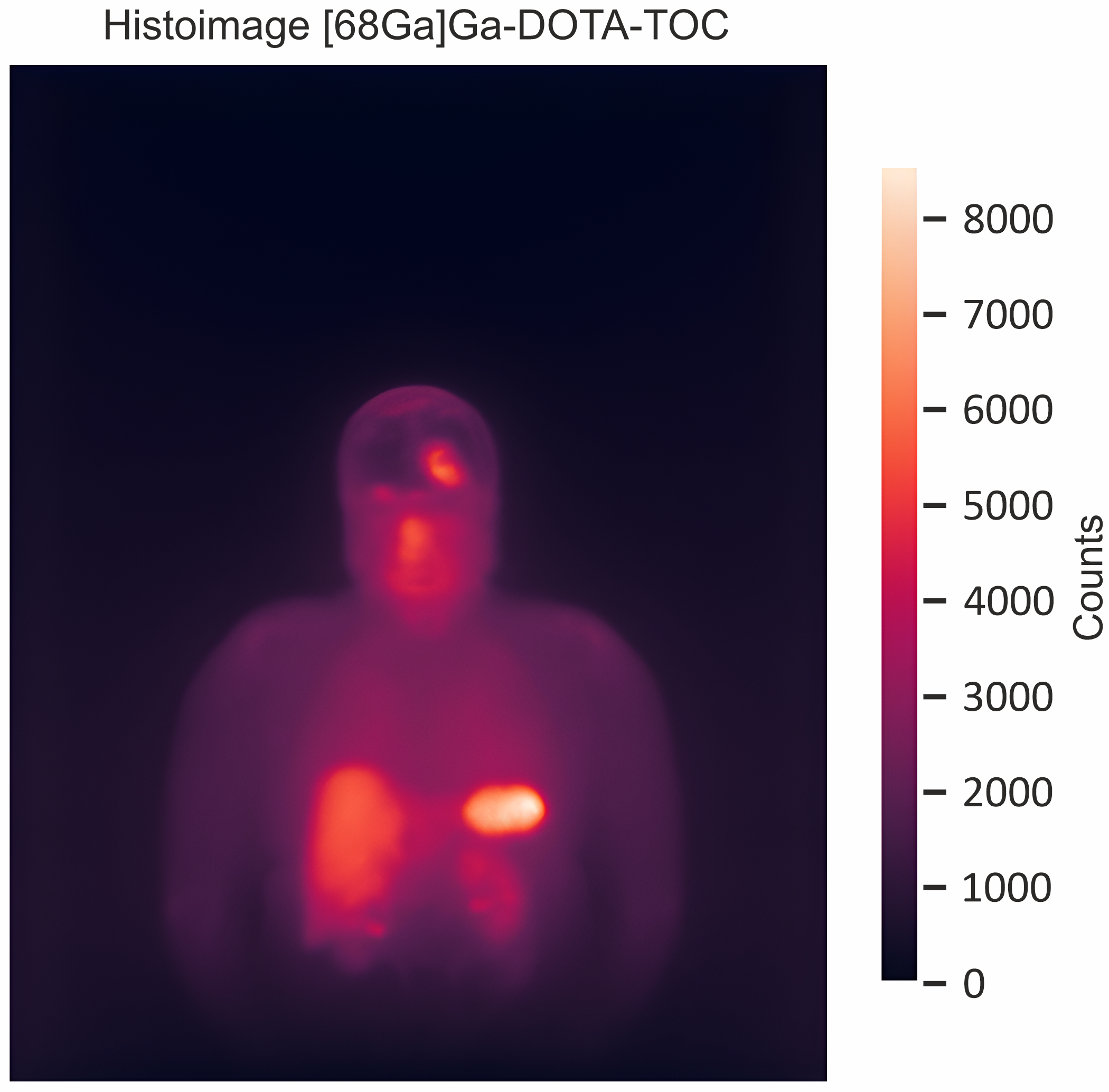}
	\caption{
		Image of a meningioma patient who underwent somatostatin receptor (SSTR)-targeting PET using ${}^{68}$Ga as a diagnostic isotope.
		The scan was acquired on a Siemens Biograph Vision Quadra system and displays the frequency of triplet events,
		i.e., the occurrences of two annihilation photons and a prompt gamma.
		Future work is aimed at deciphering the positron lifetime by measuring the time interval between the decay and annihilation photons.
	}
	\label{fig:quadra}
\end{figure}

Another radioisotope of interest for positronium lifetime imaging is $^{68}$Ga, which is routinely employed in nuclear medicine theranostics. In contrast to the above-mentioned iodine imaging, $^{68}$Ga is used with ligands that target specific overexpressed receptors on cancer cells. One example is the prostate-specific membrane antigen (PSMA), which can be both used for diagnostic purposes with isotopes like $^{68}$Ga, or for therapy with $^{177}$Lu~\cite{Alberts2021, Seifert2021a}. PSMA-therapy with $^{177}$Lu was recently approved by the FDA and EMA for advanced stages of prostate cancer, after castration resistances and several lies of therapy, including chemotherapy~\cite{Kratochwil2023}. Still, because of the low toxicity profile, especially compared with chemotherapy, earlier applications of PSMA-therapy are currently evaluated. To select patients for the treatment, PSMA-PET scans are performed, which enables the non-invasive quantification of PSMA-expression. Therefore, only candidates with strong PSMA-expression are referred to $^{177}$Lu-PSMA-therapy. Still, a fraction of approximately 30\% of patients do not sufficiently respond to the therapy~\cite{Sartor2021}. Different strategies have been proposed to refine the selection of therapy candidates, which comprise the measurement of the total tumor volume, the overall PSMA uptake in all lesions, or the assessment of the metabolic activity with [$^{18}$F]Fluorodeoxyglucose PET~\cite{Seifert2020, Seifert2021b, Seifert2023}. Preliminary evidence suggests that tumor hypoxia can also contribute to more aggressive phenotypes of prostate cancer, although this has not been systematically studied in the context of PSMA therapy~\cite{Cameron2023}. It is currently unclear, if tumor hypoxia can be investigated with positronium lifetime imaging in prostate cancer. Given the high prevalence of prostate cancer and the costs associated with PSMA-therapy, future trials on clinical positronium lifetime scans seem warranted. Similar principles also apply to other nuclear medicine treatments, like somatostatin receptor (SSTR) targeted theranostics. In analogy to PSMA, SSTR tracers can be used with $^{68}$Ga for diagnostics or with $^{177}$Lu for treatment~\cite{Pencharz2018}. One example of such a disease is meningioma, which strongly express SSTR and therefore can be targeted (Fig.~\ref{fig:quadra}). Fig.~\ref{fig:quadra} shows the first \textit{in vivo} $\beta^+ \gamma$ image of human determined with the commercial PET scanner. It demonstrates that the sensitivity of current long axial-field-of view PET scanner is sufficient to perform triple coincidence images for $2\gamma + \gamma_{\text{prompt}}$ events even when using $^{68}$Ga radionuclide of very low fraction of prompt gamma emitted in coincidence with annihilation photons with $Y_{\beta^+ + \gamma}/Y_{\beta^+} = 1.34\%$ only (Table~\ref{tab:isotopes_properties}).

With the advent of LAFOV clinical scanners, the possibility of imaging positron lifetime is rapidly approaching clinical viability. These scanners have the potential to revolutionize diagnostic imaging by their in the range of 20-fold higher sensitivity~\cite{Alberts2023}. However, significant foundational work remains to be done. Key challenges include not only capturing the occurrence of triple coincidence events but also precisely measuring the time interval between prompt gamma emission and the end of positronium life. Overcoming these technical hurdles is mandatory for the accurate imaging of positron lifetimes. The potential benefits of this technology for patient care are promising. For instance, the use of dedicated tumor hypoxia tracers, such as Fluoromisonidazole, highlights the need to utilize information in addition to molecular marker expression for treatment decisions~\cite{Xu2017}. By enabling more precise detection of hypoxic metastases, clinicians can better tailor therapies to individual patients, potentially leading to improved outcomes. Thus, the integration of positron lifetime imaging into clinical practice holds great promise for advancing patient care and improving clinical outcomes.

\section{Sensitivity of positronium imaging with \\long axial field of view PET systems}

\begin{figure}[htbp]
	\centering
	\includegraphics[width=\linewidth]{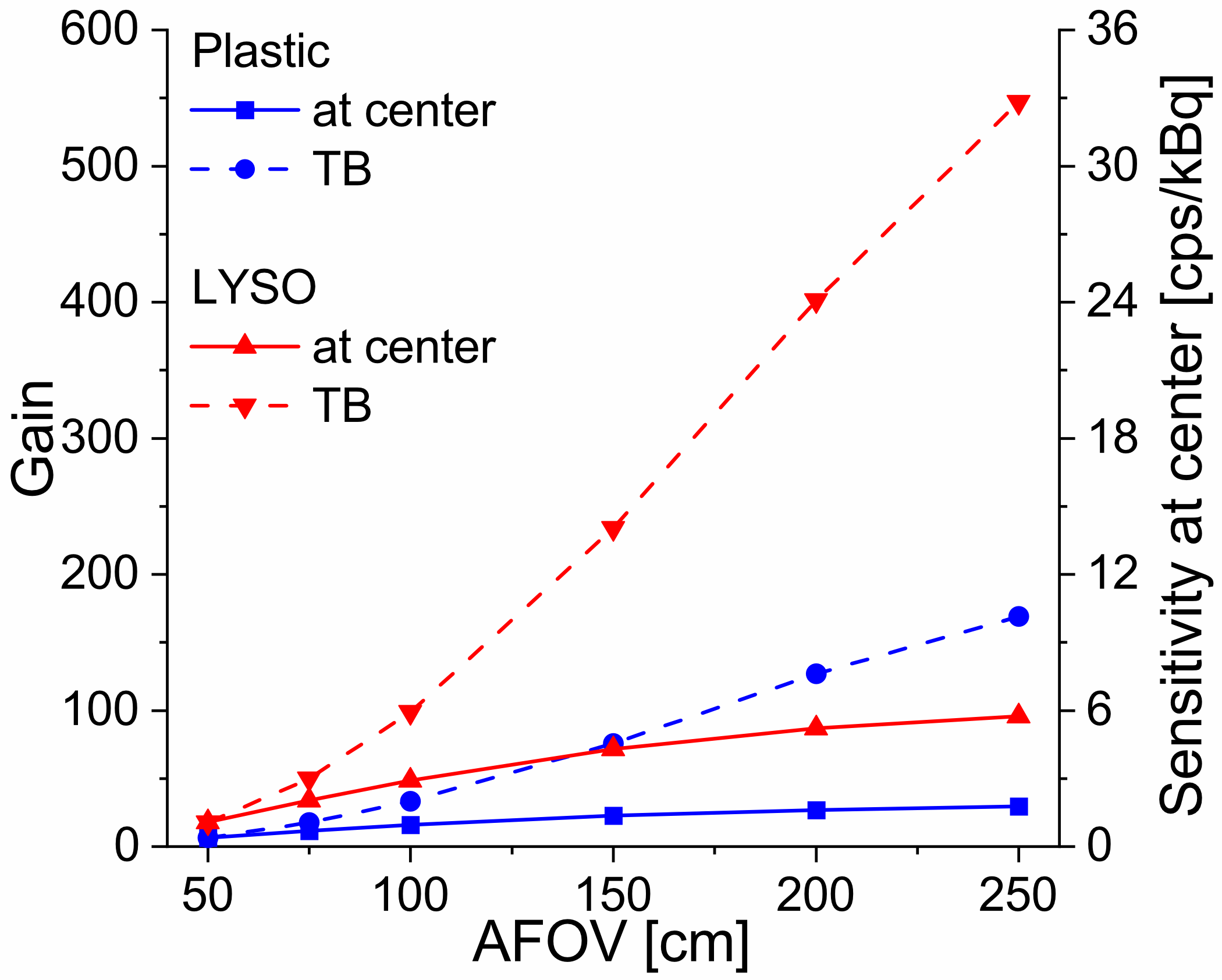}
	\caption{
		Gain of positronium imaging sensitivity as a function of the axial field of view (AFOV).
		Gain is defined as the sensitivity for positronium imaging normalized to the sensitivity of the modular J-PET scanner used to demonstrate the first \textit{in vivo} positronium images~\cite{Moskal2024a}, which is equal to 0.06~cps/kBq at 50~cm AFOV.
		Solid lines show gain at the center of the scanner, while dashed lines represent whole-body imaging.
		Blue lines correspond to plastic-based J-PET tomographs, and red lines depict LYSO crystal-based PET systems (modeled after uEXPLORER~\cite{Spencer2021}).
		Simulations were performed using the Geant4 Application for Tomographic Emission (GATE) software version~9.0~\cite{Sarrut2021}, extended by the J-PET group~\cite{Moskal2024a}.
		The right vertical axis indicates the nominal sensitivity at the center of the scanner.
		Figure adapted from~\cite{Moskal2024a}. \textcopyright~2024 American Association for the Advancement of Science. Reprinted with permission under a Creative Commons Attribution License 4.0 (CC BY).
	}
	\label{fig:gain_afov}
\end{figure}

The figure below (Fig.~\ref{fig:gain_afov}) presents the relationship between the axial field of view (AFOV) and the gain in sensitivity of positronium imaging in comparison to the sensitivity of the modular J-PET tomograph, which was used to demonstrate the first \textit{in vivo} positronium imaging~\cite{Moskal2024a}. The gain is plotted for two different cases: the sensitivity at the center (represented by solid lines) and the whole-body positronium imaging (denoted by dashed lines). The graph reveals how these two conditions perform across a range of AFOVs. The data for plastic scintillator-based J-PET tomographs are shown in blue, whereas red lines depict the sensitivity of standard LYSO crystal-based PET systems, modeled after the uEXPLORER system~\cite{Spencer2021}.

It can be observed that the LYSO-based PET scanners exhibit significantly higher gains in sensitivity, particularly as the AFOV increases, with an estimated gain nearing 600 for whole-body imaging at an AFOV of 250~cm. The plastic-based J-PET system, while showing a lower gain overall, maintains a consistent increase with a modest slope, reaching a gain of about 180 at the same AFOV. Additionally, the nominal values of the sensitivity at the center are indicated on the right axis of the graph. The LYSO-based systems achieve substantially higher sensitivity at the center compared to the plastic-based J-PET systems, highlighting the superior performance of LYSO crystals in terms of photon detection efficiency and overall system sensitivity. However, the plastic-based J-PET tomographs offer distinct advantages in other areas, such as their cost-effectiveness~\cite{Moskal2020b,Moskal2024b} and the capability to perform imaging of the degree of the entanglement~\cite{Moskal2021a,Moskal2025}.

Fig.~\ref{fig:gain_afov} indicates that sensitivity for positronium lifetime imaging with crystal-based total-body PET systems can be enhanced even more than factor of 100 compared the modular J-PET and therefore the total-body PET system and application of the $^{44}$Sc radionuclide instead $^{68}$Ga can enable positronium imaging with few thousands of times greater statistics compared to the first positronium images presented in Fig.~\ref{fig:pet_images}.

\section{Status and challenges of positronium imaging}

Electron-positron annihilation in the human body during positron emission tomography proceeds in about 40\% cases \textit{via} formation of positronium atoms~\cite{Moskal2019c,Bass2023}. In this review we presented a status of development of recently introduced method of positronium imaging~\cite{Moskal2019a} that enables imaging of parameters such as ortho-positronium lifetime ($\tau_\text{oPs}$), mean positron annihilation lifetime ($\tau_{e^+}$), mean lifetime of direct positron annihilation ($\tau_d$), the 3$\gamma$ to 2$\gamma$ annihilation rate ratio ($R_{3/2}$) and the probability of positronium formation ($P$). Parameters, that may in principle inform us about the tissue pathology at its early stage of molecular alterations, before the occurrence of functional and morphological changes, and about the degree of hypoxia~\cite{Moskal2019c,Bass2023,Moskal2021b}.

Thus far the first \textit{in vivo} in-human positronium lifetime images were demonstrated using the $^{68}$Ga radionuclide and the dedicated multi-photon J-PET scanner~\cite{Moskal2024a}. The obtained images mark a significant milestone in the development of positronium imaging indicating \textit{in vivo} differences in $\tau_\text{oPs}$ and $\tau_{e^+}$ between healthy brain tissues and glioblastoma tissues, though still with the very low statistics. Much higher statistics was achieved for the following first \textit{in vivo} positronium lifetime measurements of humans performed with long axial-field-of-view Biograph Vision Quadra commercial PET scanner using $^{68}$Ga labelled pharmaceuticals and $^{82}$Rb-Chloride~\cite{Mercolli2024}, thus marking another milestone in translating the positronium imaging into clinics. Notably, the measured $\tau_\text{oPs}$ in the right heart ventricle was higher than in the left heart ventricle indicating a possible shortening of the lifetime in more oxygenated blood~\cite{Mercolli2024}. In addition, the positronium lifetime measurements performed with clinical scanners for phantoms and certified materials were reported also for studies with clinically available radionuclides as $^{68}$Ga and $^{82}$Rb~\cite{Steinberger2024,Huang2024a}, $^{124}$I~\cite{Steinberger2024,Mercolli2025a} and also for $^{44}$Sc~\cite{Mercolli2025b,Das2025}. In the studies of positronium lifetime imaging with $^{82}$Rb at PennPET Explorer an application of iterative time-thresholding reconstruction was also demonstrated for the extended phantoms~\cite{Huang2024a}.

The so far reported studies of positronium lifetime imaging demonstrated the feasibility of application of this method in commercial PET scanners in which a triggerless data acquisition mode can be adopted~\cite{Steinberger2024,Huang2024a,Mercolli2025a}. The first studies with commercial systems revealed that sensitivity of long axial field-of-view scanners such as e.g.\ Biograph Vision Quadra or PennPET Explorer will be sufficient for effective positronium imaging. See for example triple coincidence image shown in Fig.~\ref{fig:quadra} that was acquired with application of $^{68}$Ga having only 1.34\% of $\beta^+$+$\gamma$ events with positron and prompt gamma. Application of $^{44}$Sc would increase it by factor of $\sim$75. Scandium is one of the most promising radionuclides for positronium imaging~\cite{Moskal2020b,Das2023}. However, for its clinical applications in human the effort is required for its FDA approval of $^{44}$Sc labelled pharmaceuticals. Moreover, presently the commercial scanners as e.g.\ Biograph Vision Quadra collect all signals larger than 726~keV in a single bin being unable to resolve photoelectric interaction of prompt gamma from $^{44}$Sc having energy of 1157~keV~\cite{Mercolli2025b}. Furthermore, the iterative positronium lifetime imaging methods need to be tested on \textit{in vivo} data with humans and further developed to account for scattering and attenuation of photons in the patient.

Once the commercial scanners are modernized with the ability to simultaneously record annihilation and deexcitation photons in required energy range, and once the iterative reconstruction will provide spatial resolution of positronium images comparable with the present PET images, the studies will concentrate on finding the diagnostic meaning of positronium lifetime and developing the robust effective way of extracting the clinically useful parameters being either discussed $\tau_\text{oPs}$, $\tau_{e^+}$, $\tau_d$, $R_{3/2}$, and $P$ or perhaps their combination.

\section*{Acknowledgments}

This work was supported by the National Science Centre of Poland through grants MAESTRO no. 2021/42/A/ST2/00423, OPUS no. 2021/43/B/ST2/02150 and OPUS-LAP no. 2022/47/I/NZ7/03112 as well as by the SciMat and qLife Priority Research Area budgets under the programme Excellence Initiative- Research University at the Jagiellonian University.

\vspace{1em}

The authors declare the following financial interests/personal relationships which may be considered as potential competing interests with the work reported in this paper: P.M. is an inventor on a patent related to this work. Patent nos.: (Poland) PL 227658, (Europe) EP 3039453, and (United States) US 9,851,456], filed (Poland) 30 August 2013, (Europe) 29 August 2014, and (United States) 29 August 2014; published (Poland) 23 January 2018, (Europe) 29 April 2020, and (United States) 26 December 2017. A.R. has received research support and speaker honoraria from Siemens. K.S. received research grants from Novartis and Siemens, further conference sponsorships from United Imaging, Siemens, and Subtle Medical outside of the submitted work. R.S. has received research/travel support from Boehringer Ingelheim Fund and Else Kröner-Fresenius-Stiftung, as well as travel support and lecture fees from Novartis and Boston Scientific, outside the submitted work. W.S. is employee of Siemens Medical Solutions USA. Other authors declare that they have no known conflicts of interest in terms of competing financial interests or personal relationships that could have an influence or are relevant to the work reported in this paper.

\bibliographystyle{IEEEtran}

\vfill

\end{document}